\documentstyle[12pt,epsfig]{article}
\catcode`\@=11
  
\def\@normalsize{\@setsize\normalsize{15pt}\xiipt\@xiipt
\abovedisplayskip 14pt plus3pt minus3pt%

\belowdisplayskip \abovedisplayskip
\abovedisplayshortskip  \z@ plus3pt%
\belowdisplayshortskip  7pt plus3.5pt minus0pt}
\def\small{\@setsize\small{13.6pt}\xipt\@xipt
\abovedisplayskip 13pt plus3pt minus3pt%
\belowdisplayskip \abovedisplayskip
\abovedisplayshortskip  \z@ plus3pt%
\belowdisplayshortskip  7pt plus3.5pt minus0pt
\def\@listi{\parsep 4.5pt plus 2pt minus 1pt
            \itemsep \parsep
            \topsep 9pt plus 3pt minus 3pt}}
 
\def\underline#1{\relax\ifmmode\@@underline#1\else
        $\@@underline{\hbox{#1}}$\relax\fi}
\@twosidetrue
\relax
 
\catcode`@=12
 
\catcode`\@=11

\def\section{\@startsection{section}{1}{\z@}{3.5ex plus 1ex minus
   .2ex}{2.3ex plus .2ex}{\large\bf}}

\def\ps@headings{\def\@oddfoot{}\def\@evenfoot{}
\def\@oddhead{\hbox{}\hfill
        \makebox[.5\textwidth]{\raggedright\ignorespaces --\thepage{}--
        \hfill }}
\def\@evenhead{\@oddhead}
\def\subsectionmark##1{\markboth{##1}{}}
}
 
\ps@headings
 
\catcode`\@=12
 
\relax

\def\figcap{\section*{Figure Captions\markboth
        {FIGURECAPTIONS}{FIGURECAPTIONS}}\list
        {Fig. \arabic{enumi}:\hfill}{\settowidth\labelwidth{Fig. 999:}
        \leftmargin\labelwidth
        \advance\leftmargin\labelsep\usecounter{enumi}}}
 \relax
\def\tablecap{\section*{Table Captions\markboth
        {TABLECAPTIONS}{TABLECAPTIONS}}\list
        {Table \arabic{enumi}:\hfill}{\settowidth\labelwidth{Table 999:}
        \leftmargin\labelwidth
        \advance\leftmargin\labelsep\usecounter{enumi}}}
 \relax
\def\reflist{\section*{References\markboth
        {REFLIST}{REFLIST}}\list
        {[\arabic{enumi}]\hfill}{\settowidth\labelwidth{[999]}
        \leftmargin\labelwidth
        \advance\leftmargin\labelsep\usecounter{enumi}}}
 \relax
 
\catcode`\@=11
 
\def\marginnote#1{}
\newcount\hour
\newcount\minute
\newtoks\amorpm
\hour=\time\divide\hour by60
\minute=\time{\multiply\hour by60 \global\advance\minute by-
\hour}
\edef\standardtime{{\ifnum\hour<12 \global\amorpm={am}%
    \else\global\amorpm={pm}\advance\hour by-12 \fi
    \ifnum\hour=0 \hour=12 \fi
    \number\hour:\ifnum\minute<100\fi\number\minute\the\amorpm}}
\edef\militarytime{\number\hour:\ifnum\minute<100\fi\number\minute}
\def\draftlabel#1{{\@bsphack\if@filesw {\let\thepage\relax
  \xdef\@gtempa{\write\@auxout{\string
    \newlabel{#1}{{\@currentlabel}{\thepage}}}}}\@gtempa
    \if@nobreak \ifvmode\nobreak\fi\fi\fi\@esphack}
     \gdef\@eqnlabel{#1}}
\def\@eqnlabel{}
\def\@vacuum{}
\def\draftmarginnote#1{\marginpar{\raggedright\scriptsize\tt#1}}
\def\draft{\oddsidemargin -.5truein
        \def\@oddfoot{\sl preliminary draft \hfil
        \rm\thepage\hfil\sl\today\quad\militarytime}
        \let\@evenfoot\@oddfoot \overfullrule 3pt
        \let\label=\draftlabel
        \let\marginnote=\draftmarginnote
 
\def\@eqnnum{(\theequation)\rlap{\kern\marginparsep\tt\@eqnlabel}%
\global\let\@eqnlabel\@vacuum}  }
\def\preprint{\twocolumn\sloppy\flushbottom\parindent 1em
        \leftmargini 2em\leftmarginv .5em\leftmarginvi .5em
        \oddsidemargin -.5in    \evensidemargin -.5in
        \columnsep 15mm \footheight 0pt
        \textwidth 250mmin      \topmargin  -.4in
        \headheight 12pt \topskip .4in
        \textheight 175mm
        \footskip 0pt
 
\def\@oddhead{\thepage\hfil\addtocounter{page}{1}\thepage}
        \let\@evenhead\@oddhead \def\@oddfoot{} \def\@evenfoot{}
}
\def\titlepage{\@restonecolfalse\if@twocolumn\@restonecoltrue\onecolumn
     \else \newpage \fi \thispagestyle{empty}\c@page\z@
        \def\thefootnote{\fnsymbol{footnote}} }
\def\endtitlepage{\if@restonecol\twocolumn \else  \fi
        \def\thefootnote{\arabic{footnote}}
        \setcounter{footnote}{0}}  
\catcode`@=12
\relax
 
\def\ps@headings{\def\@oddfoot{}\def\@evenfoot{}
\def\@oddhead{\hbox{}\hfill
        \makebox[.5\textwidth]{\raggedright\ignorespaces --\thepage{}--
        \hfill }}
\def\@evenhead{\@oddhead}
\def\subsectionmark##1{\markboth{##1}{}}
}
 
\ps@headings
 
\relax

\newcommand{\newc}{\newcommand}
\newc{\ra}{\rightarrow}
\newc{\lra}{\leftrightarrow}
\newc{\beq}{\begin{equation}}
\newc{\eeq}{\end{equation}}
\newc{\bea}{\begin{eqnarray}}
\newc{\eea}{\end{eqnarray}}

\def\la{\lambda}

\newc{\sm}{Standard Model}
\newc{\smd}{Standard Model}
\newc{\barr}{\begin{eqnarray}}
 \newc{\earr}{\end{eqnarray}}

\ps@headings
 
\relax
 
\def\firstpage#1#2#3#4#5#6{
\begin{document}

\begin{titlepage}
\nopagebreak
\title{\begin{flushright}
        \vspace*{-0.8in}
{ \normalsize  hep-ph/9911459 \\
ACT-12/99 \\
CERN-TH/99-237 \\
CTP-TAMU-45/99 \\
FISIST/22-99/CFIF \\
}
\end{flushright}
\vfill
{#3}}
\author{\large #4 \\[0.7cm] #5}
\maketitle
\vskip -7mm
\nopagebreak
\begin{abstract}
{\noindent #6}
\end{abstract}
\vfill
November 1999  
\begin{flushleft}
\rule{16.1cm}{0.2mm}\\[-3mm]


\end{flushleft}
\thispagestyle{empty}
\end{titlepage}}
 
\def\simlt{\stackrel{<}{{}_\sim}}
\def\simgt{\stackrel{>}{{}_\sim}}
\date{}
\firstpage{3118}{IC/95/34}
{\Large \bf 
Charged-Lepton-Flavour Violation in the Light
of the Super-Kamiokande Data
}
{\bf \normalsize \bf John Ellis$^{\,a}$,  M.E. G\'omez$^{\,b}$,
G.K. Leontaris$^{\,a,c}$,
S. Lola$^{\,a}$ ${\rm and}$ D.V. Nanopoulos$^{\,d}$}
{\normalsize\sl
$^a$Theory Division, CERN, CH 1211 Geneva 23, Switzerland\\[2.5mm]
\normalsize\sl
$^b$Centro de F\'{\i}sica das Interac\c{c}\~{o}es Fundamentais
(CFIF), Departamento de F\'{\i}sica,  \\ 
\normalsize\sl
Instituto Superior T\'ecnico,  Av. Rovisco Pais,  1049-001 Lisboa,
Portugal\\[2.5mm]
\normalsize\sl
$^c$Theoretical Physics Division, Ioannina University,
GR-45110 Ioannina, Greece\\[2.5mm]
\normalsize\sl
$^d$Center for Theoretical Physics, Department of Physics,\\[-1.0mm]
\normalsize\sl
  Texas A\&M
 University, College Station, TX 77843 4242,  USA;\\[-1.0mm]
\normalsize\sl
Astroparticle Physics Group, Houston Advanced Research Center
(HARC),
\\[-1.0mm]
\normalsize\sl
The Mitchell Campus, Woodlands, TX 77381, USA;\\[-1.0mm]
\normalsize\sl
Academy of Athens, Chair of Theoretical Physics, Division of Natural
Sciences,\\[-1.0mm]
\normalsize\sl
 28 Panepistimiou Ave., Athens GR-10679,  Greece}
{Motivated by the data from Super-Kamiokande and elsewhere
indicating oscillations of atmospheric and solar neutrinos, we 
study charged-lepton-flavour violation,
in particular the radiative decays $\mu \rightarrow
e \gamma$ and $\tau \rightarrow \mu \gamma$, but also commenting
on $\mu \ra 3e$ and $\tau \to 3 \mu/e$ decays, as well as
$\mu \to e$ conversion on nuclei.
We first show how the renormalization group may be used to calculate
flavour-violating soft supersymmetry-breaking masses for charged sleptons
and sneutrinos in models with universal input parameters. 
Subsequently, we classify possible patterns of lepton-flavour violation
in the context of phenomenological neutrino mass textures
that accommodate the Super-Kamiokande data, giving examples based on
Abelian flavour symmetries.
Then we calculate in these examples rates for $\mu \rightarrow
e \gamma$ and $\tau \rightarrow \mu \gamma$, which may be close
to the present experimental upper limits, and show how they 
may distinguish between the different generic mixing patterns.
The rates are promisingly large when the soft supersymmetry-breaking
mass parameters are chosen to be consistent with
the cosmological relic-density constraints.
In addition, we discuss  
$\mu \rightarrow e$ conversion on Titanium, which may also
be accessible to future experiments.}
 
\pagebreak

\setcounter{page}{1}

\section{Introduction}

There has been increasing interest in massive neutrinos
during the past year, triggered principally by 
the Super-Kamiokande data~\cite{SKam} 
on the $\nu _{\mu }/\nu _{e }$ ratio in the atmosphere.
The latter was found to be significantly
smaller than the Standard Model expectations,
with a characteristic azimuthal-angle dependence
indicating the presence of neutrino oscillations.
The data analysis favours
$\nu_{\mu } \to \nu_{\tau }$ oscillations,
with parameters in the ranges
\begin{eqnarray}
\delta m_{\nu _{\mu }\nu _{\tau }}^{2} &\sim &(10^{-2}\;{\rm to}%
\;10^{-3})\;{\rm eV^{2}} \\
\sin^{2}2\theta _{\mu \tau } &\geq &0.8  \label{atmos}
\end{eqnarray}
Dominance by $\nu _{\mu }\rightarrow \nu _{e}$
oscillations is disfavoured by the 
Super-Kamiokande data on electron-like events~\cite{SKam}, 
as well as by the data from the Chooz reactor experiment~\cite{chooz}.
Oscillations involving a sterile neutrino
are disfavoured, but not yet excluded, by a detailed study of
the azimuthal-angle dependence of muon-like events~\cite{SKam}
and by measurements of $\pi^0$ production.
Moreover, in most theoretical models sterile neutrinos
tend to be heavy. Therefore, we consider $\nu_\mu \ra \nu_\tau$
as the `established' hypothesis for the atmospheric neutrino
data.

In addition, the long-standing deficit of solar $\nu_e$ measured on Earth
may also be explained via neutrino oscillations,
either {\it in vacuo} or enhanced in matter by the
Mikheyev-Smirnov-Wolfenstein (MSW) mechanism~\cite{MSW}.
The first option would require 
$\delta m_{\nu _{e}\nu _{\alpha }}^{2} 
\sim (0.5-1.1)\times 10^{-10}~{\rm eV}^2$,
where $\alpha $ is $\mu $ or $\tau$. 
MSW oscillations~\cite{MSW}, on the other hand, require
$ \delta m_{\nu _{e}\nu _{\alpha }}^{2} \sim (0.3-20) \times 10^{-5}
~{\rm eV}^2$ with either large $\sin^{2}2\theta _{e \alpha } \sim 1$
or small $\sin^{2}2\theta _{e \alpha} \sim 10^{-2}$.
The presence of either $\nu_e \ra \nu_\mu$ or $\nu_e \ra \nu_\tau$
oscillations at a high level is, therefore, an open question.

Both the solar and atmospheric neutrino
data can be accommodated in a natural way
in schemes with three light
neutrinos with at least one large mixing angle and hierarchical
masses, of the order of the required
mass differences:
$m_{3} \sim (10^{-1} \; {\rm to} \;
10^{-1.5})$ eV
and $m_{2} \sim (10^{-2} \; {\rm to} \;
10^{-3})$ eV $\gg m_3$.
On the other hand,
if neutrinos were also to provide significant hot dark matter,
three almost-degenerate
neutrinos with  masses of $\approx 1$ eV
would be needed.

Neutrino oscillations involve violations of the
individual lepton numbers $L_{e, \mu, \tau}$,
raising the prospect that there might also exist
observable processes that violate charged-lepton
number conservation \cite{neutrinoLFVns,
neutrinoLFVs}, such as 
 $\mu \rightarrow e \gamma, 3 e$ and
$\tau \rightarrow \mu \gamma, 3\mu/e$, and $\mu \to e$
conversion on heavy nuclei~\cite{neutrinoLFVs,oldflav,a,gllv,FNS,KO}.
We recall briefly the present experimental upper limits on the most
interesting of these decays for our subsequent discussion:
\begin{eqnarray}
BR(\mu \ra e \gamma) \, < \, 1.2 \times 10^{-11} \, &:& \, \,
\cite{Brooks} \\
BR(\mu^+ \ra e^+ e^+ e^-) \, < \, 1.0 \times 10^{-12} \, &:& \, \,
\cite{Bellgardt} \\
R(\mu^- Ti \ra e^- Ti) \, < \, 6.1 \times 10^{-13} \, &:& \, \,
\cite{Wintz} \\
BR(\tau \ra \mu \gamma) \, < 1.1 \times \, 10^{-6} \, &:& \, \,
\cite{CLEO}
\end{eqnarray}
Projects are currently underway to improve several of these upper limits
significantly:
\begin{eqnarray}
BR(\mu \ra e \gamma) \, \ra \, 10^{-14} \, &:& \, \,
\cite{Barkov} \\
R(\mu^- Ti \ra e^- Ti) \, \ra \, {\rm few} \times 10^{-14} \, &:& \, \,
\cite{SINDRUM} \\
R(\mu^- Al \ra e^- Al) \, \ra \, 1 \times 10^{-16} \, &:& \, \,
\cite{MECO} \\
BR(\tau \ra \mu \gamma) \, \ra 1 \times \, 10^{-9}~? \, &:& \, \,
\cite{LHC}
\end{eqnarray}
and there are active discussions of intense $\mu$ sources that might
enable the upper limits on $\mu \ra e$ transitions to be improved
by several further orders of magnitude~\cite{mufact}.

We evaluate the
possibility of charged-lepton-flavour violation using the most natural
mechanism for obtaining neutrino masses
in the sub-eV range, namely the see-saw mechanism~\cite{seesaw}, which
involves Dirac neutrino masses $m_{\nu}^D$ of the same order as
the charged-lepton and quark masses, and heavy Majorana
masses $M_{\nu_R}$, leading to light effective neutrino mass
matrices:
\begin{equation}
m_{eff}
=m^D_{\nu}\cdot (M_{\nu_R})^{-1}\cdot m^{D^{\normalsize T}}_{\nu}.
\label{eq:meff}
\end{equation}
Neutrino-flavour mixing may then occur through either the
Dirac and/or the Majorana mass matrices, which may also
feed flavour violation through to the charged leptons.

The specific mechanism explored in this paper is
renormalization of the sneutrino and slepton masses
in a supersymmetric theory via the neutrino Dirac couplings
$\lambda^D_\nu$ \cite{neutrinoLFVs}. It is well known that the
prototypical charged-lepton flavour-violating process
$\mu \to e \gamma$ provides one of the most stringent upper
limits on flavour violation in the Minimal Supersymmetric
extension of the Standard Model (MSSM). If the soft
supersymmetry-breaking sneutrino and slepton masses
were non-universal before renormalization, it would be very
difficult to understand why this decay was not seen long ago.
Universal scalar masses arise naturally in
no-scale supergravity models~\cite{Ellis:1984bm}, 
the framework favoured here, as well as in
gauge-mediated models~\cite{gaugmed}.
In the universal supergravity case, the soft
supersymmetry-breaking sneutrino and slepton masses
are subject to calculable and non-trivial renormalization via
Dirac neutrino couplings at scales between $M_{GUT} \sim 10^{16}$~GeV
and $M_{\nu_R} \sim 10^{13}$~GeV.

The predictions of this class of universal supergravity models
are quite characteristic. In non-supersymmetric models with
massive neutrinos, the amplitudes for the
charged-lepton-flavour violation are 
proportional to inverse powers of
the right-handed neutrino mass scale 
$M_{\nu_R}$~\cite{neutrinoLFVns}. Since the latter
is much higher than the electroweak scale, the rates 
for rare decays such as $\mu \to e \gamma$ are  extremely 
suppressed~\cite{neutrinoLFVns}.
On the other hand, in supersymmetric models these
processes are only suppressed
by inverse powers of the supersymmetry breaking scale, which is 
at most $1$~TeV \cite{neutrinoLFVs}.
Among such models, those with non-universal
input scalar masses at the GUT scale generally predict excessive
rates for rare charged-lepton-flavour violation, whereas 
they are very suppressed in no-scale~\cite{Ellis:1984bm}
and gauge-mediated models~\cite{gaugmed}. The class of
supergravity models with universal scalar masses that we consider here
toe the fine line between excessive and unobservable
charged-lepton-flavour violation, as we discuss in more detail below.

Here we re-analyze the prospects for
charged-lepton-flavour violation in this theoretical framework, 
using as a guide the indications from Super-Kamiokande and elsewhere
on neutrino masses and mixing. Within the general see-saw
scenario, 
solutions with various differences in the neutrino mass-matrix
structure have been proposed,
including models with
maximal~\cite{guid,ELLN2}, close-to-maximal~\cite{GGR,GGR2} 
and bi-maximal~\cite{bimaximal} neutrino mixing.
In the following, we categorize models according to whether
the off-diagonal elements in
their Dirac and Majorana couplings `match' in such a way that
their mixing is almost two-generational (and hence predominantly in the
$\mu - \tau$ sector), and more general models
in which they `mismatch', and the mixing is essentially
three-generational (and substantial also in the $\mu - e$ and $\tau - e$
sectors). We provide examples of the `matched' and
`mismatched' categories in schemes
with Abelian~\cite{abel,GGR,GGR2} flavour symmetries, and comment
on the possibilities with
non-Abelian~\cite{nonabel} flavour symmetries, as well as in a
string-derived flipped $SU(5)$ model~\cite{aehn},
whose characteristic property is the appearance of large off-diagonal 
$\mu - e$ couplings in the Dirac neutrino-mass matrix~\cite{ELLN2}.
As we exemplify with calculations in the Abelian models, the rates for the
radiative decays $\mu \to e \gamma$ and
$\tau \to \mu \gamma$ may offer good prospects for testing
different textures. These decays may well take place at observable
rates, and different neutrino-mass models
correlate their decay rates in characteristically different
ways, enabling $\mu \to e \gamma$ and
$\tau \to \mu \gamma$ to serve as useful
diagnostic tools for neutrino-mass models.
Finally, we calculate in some models the rate for
$\mu \to e$ conversion on Titanium, which may also be
accessible to future experiments, and comment briefly on
$\mu \to 3 e$ decay.

\section{General Aspects of Charged-Lepton-Flavour Violation in
Supersymmetric Models with Universal Breaking}

We first display in Fig.~1 the one-loop diagrams 
that give rise to $\mu \rightarrow 
e\gamma$, noting that the $\tau\ra \mu\gamma$-decay is
generated by an analogous set of graphs.
We later extend the discussion to include $\mu \ra 3 e$ decay,
$\tau \ra 3 \mu/e$ and $\mu \ra e$ conversion.
The matrix element of the
electromagnetic-current operator between two distinct lepton  states $l_i$
and $l_j$ is given in general by
\begin{eqnarray}
{\cal T}_\la &=& \langle l_i|(p-q)|{\cal J}_\la|l_j(p)\rangle\nonumber\\
{  }&=&{\bar u_i}(p-q)
      \{ m_j i\sigma_{\la\beta}q^\beta
               \left(A^L_MP_L+A^R_MP_R\right)+\nonumber\\
{  }&&\quad\quad    (q^2\gamma_\la-q_\la\gamma\cdot q)
               \left(A^L_EP_L+A^R_EP_R\right)
      \} u_j(p)
\label{general}
\end{eqnarray}
where $q$ is the photon momentum. The 
coefficients $A_M$ and $A_E$ receive
contributions from both neutralino (n)/charged-slepton (Fig.~1(a))
and
chargino (c)/sneutrino (Fig.~1(b) exchanges:
\begin{equation}
A_M^{L,R}=A_{M(n)}^{L,R}+A_{M(c)}^{L,R},\quad
A_E^{L,R}=A_{E(n)}^{L,R}+A_{E(c)}^{L,R}
\label{ampl}
\end{equation}
The amplitude of the process is then proportional to ${\cal T}_\la
\epsilon^\la$, where $\epsilon^\la$ is the photon-polarization vector.

\begin{figure}[h]
\begin{center}
\epsfig{file=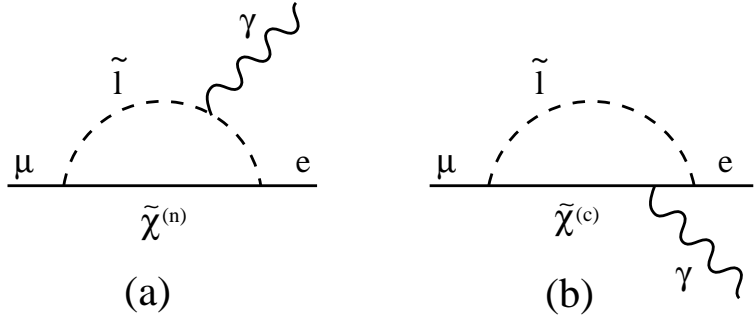,width=10cm}
\end{center}
\caption{{\it Generic Feynman diagrams for $\mu\ra e\gamma$
decay: $\tilde l$ represents a charged slepton (a) or sneutrino (b), and
$\tilde\chi ^{(n)}$ and $\tilde\chi ^{(c)}$ represent  
neutralinos and charginos respectively.}}
\label{figure1}
\end{figure}

We recall that the easiest way to determine the loop momentum-integral
contributions
to the coefficients $A_{M,E}$ is to identify, in the corresponding
diagram, terms
proportional to $(p\cdot\epsilon)$ and $(q\cdot\epsilon)$. The
coefficient of the former is proportional to the momentum-integral
contribution to the $\sigma_{\la\beta}$ term in (\ref{general}), while
the coefficient of the latter is proportional to the difference
between the
momentum-integral contributions to the $\sigma_{\la\beta}$ and
$(q^2\gamma_\la-q_\la\gamma\cdot q)$ terms. Defining
$x \equiv M^2/m^2$, where $M$ is the chargino (neutralino)  mass and $m$
the
sneutrino (charged slepton) mass, the following functions appear in
the $A_M$ terms \cite{KO,neutrinoLFVs}:
\begin{equation}
\begin{array}{ll}
A_{M(n)}:\quad&\frac{1}{6(1-x)^4}(1-6x+3x^2+2x^3-6x^2\log x) \quad\hbox{and}\\
        &\frac{1}{(1-x)^3}(1-x^2+2x\log x)\frac{M}{m_{l_j}}\\
A_{M(c)}:\quad&\frac{1}{6(1-x)^4}(2+3x-6x^2+x^3+6x\log x)\quad \hbox{and}\\
        &\frac{1}{(1-x)^3}(-3+4x-x^2-2\log x)\frac{M}{m_{l_j}}
\end{array}
\label{AM}
\end{equation}
where $m_{l_j}$ is the mass of the $l_j$ lepton, while for the $A_E$ 
terms we have:
\begin{equation}
\begin{array}{ll}
A_{E(n)}:\quad&\frac{1}{(1-x)^4}(2-9x+18x^2-18x^3+6x^3\log x) \\
A_{E(c)}:\quad&\frac{1}{(1-x)^4}(16-45x+36x^2-7x^3+6(2-3x)\log x).
\end{array}
\label{AE}
\end{equation}
Note in this case the lack of terms proportional to the gaugino mass $M$.
The branching ratio (BR) of the decay $l_j\ra l_i+\gamma$ is 
then given by:
\begin{equation}
BR(l_j\ra l_i\gamma)=\frac{48\pi^3\alpha}{G_F^2}
               \left((A_M^L)^2+(A_M^R)^2\right)
\label{BR}
\end{equation}
We see, therefore, that
the branching ratios for 
radiative lepton decays involve the masses
of several supersymmetric particles at low energies.

As stated in the Introduction, in this work we assume
universal scalar masses and trilinear terms $A$ at the GUT scale.
However, the physical values of these masses to be used
in~(\ref{AM},\ref{AE},\ref{BR}) have to be obtained by
integrating  the renormalization-group equations of the 
MSSM supplemented with right-handed neutrinos, found, for
example, in~\cite{rge}).
The Dirac neutrino and charged lepton Yukawa
couplings cannot, in general, be diagonalized simultaneously.
Since both these sets of lepton Yukawa
couplings appear 
in the renormalisation-group equations, the lepton Yukawa
matrices and the slepton mass matrices can not be simultaneously
diagonalized at low energies either.
Indeed, in the basis where $m_{\ell}$
is diagonal, the slepton-mass matrix acquires 
non-diagonal contributions from renormalization at scales below
$M_{GUT}$, of the form \cite{neutrinoLFVs}:
\bea
\delta\tilde{m}_{\ell}^2\propto \frac 1{16\pi^2} (3 + a^2)
\ln\frac{M_{GUT}}{M_N}\lambda_D^{\dagger} \lambda_D m_{3/2}^2,
\label{offdiagonal}
\eea
where $\lambda_D$ is the Dirac neutrino Yukawa coupling, 
$M_N$ is the intermediate scale where 
the effective neutrino-mass operator is formed, and
$a$ is  related to the trilinear mass parameter: $A_l= a m_{3/2}$,
where $m_{3/2}^2$ is the common value of the scalar masses at the GUT
scale. As a result, the diagrams of Fig.~1 lead to radiative decays
of charged leptons~\footnote{We note that a complete
renormalization-group analysis would also involve
the quark and squark 
sector, and require a treatment of supersymmetric thresholds. However,
the inclusion of such detailed effects would not affect the main
conclusions of our analysis, so we may neglect them for our purposes.}.

We obtain the physical charged-slepton masses by numerical
diagonalization of the following $6 \times 6 $ matrix:
\begin{equation}
\label{eq:66}
\tilde{m}_{\ell}^2=\left(\begin{array}{cc} m_{LL}^2&m_{LR}^2\\
                                   m_{RL}^2&m_{RR}^2 \end{array}\right)
\end{equation}
where all tha entries are $3 \times 3$ matrices in flavour space.
Using the superfield basis where $\lambda_{\ell}$ is diagonal, it is convenient 
for later use to write the $3\times 3$ entries of (\ref{eq:66}) in the form:
\begin{eqnarray}
m_{LL}^2&=& (m_{\tilde{\ell}}^\delta)^2+ \delta m_{\nu_D}^2+m_{\ell}^2 -
M_Z^2       
(\frac{1}{2} -sin^2\theta_W) cos 2\beta\\
m_{RR}^2&=& (m_{\tilde{e_R}}^\delta)^2+m_{\ell}^2
 -M_Z^2 sin^2\theta_W cos 2\beta \\
m_{RL}^2&=& (A_e^\delta +\delta A_e - \mu tan\beta) m_{\ell}\\
m_{LR}^2&=& m_{RL}^{2\dagger}
\label{bits}
\end{eqnarray}
where $\tan\beta$ is the standard ratio of the two MSSM Higgs vevs,
$(m_{\tilde{\ell}}^\delta)^2, (m_{\tilde{e_R}}^\delta)^2$ and $A_e^\delta$ 
denote the diagonal contributions to the corresponding matrices,
obtained by numerical integration of the
renormalization-group equations, and $\delta m_{\nu_D}^2$ and $\delta A_{l}$
denote the off-diagonal terms that appear because
$\lambda_D$ and $\lambda_{\ell}$ may not be diagonalized
simultaneously (\ref{offdiagonal}).

The  full mass matrix for left- and right-handed 
sneutrinos has a $12\times 12$ structure,
given in terms of $3\times 3$
Dirac,  Majorana  and sneutrino mass matrices.
The effective $3  \times 3$
mass-squared matrix for the left-handed sneutrinos has the same
form as the $m_{LL}^2$
part (\ref{bits}) of the $6 \times 6$ charged-slepton matrix
(\ref{eq:66}), with the difference that now the Dirac masses
are absent. One might have
expected that - as in the case of charged sleptons -
the Dirac terms would induce considerable mixing effects. However, we
show here that this is {\it not} the case in the sneutrino-mass matrix.
Due to the vastly different scales involved in the full $12\times 12 $
sneutrino mass matrix, a complete analysis is not straightforward.
Instead, we construct an effective $6 \times 6$ matrix
for the light sector, by applying matrix
perturbation theory, in analogy with to the see-saw
mechanism~\cite{seesaw}. To second order, the result is:
\beq
(m_{\tilde{\nu}}^2)_{eff} =
\left(
\begin{array}{cc}
m_{\tilde{\ell}}^2 +{\cal O}(m_{\tilde{\ell}}^4M^{-2})
 &
 ( (2 A_\nu + A_N) - 2 \mu \cot\beta) \cdot \\
\hspace{0.9 cm}
&
\hspace{0.9 cm}
(m_D M^{-1} m_D^\dagger) \\
  & \\
 ( (2 A_\nu + A_N) - 2 \mu \cot\beta)\cdot &
m_{\tilde{\ell}}^2 +{\cal O}(m_{\tilde{\ell}}^4M^{-2})
\\
\hspace{0.9 cm}
(m_D M^{-1} m_D^\dagger)^* &
\hspace{0.9 cm}
\end{array}
\right)
\label{sixbysix}
\eeq
The first- and second-order terms in (\ref{sixbysix}) are obtained
assuming that all the parameters
as real and the matrices $A_{\nu,N}$ are proportional to the identity.
Notice that the second-order terms along the diagonal
can be neglected, whereas the first-order off-diagonal
terms must be retained, since they lead to
complete mixing of the pair-wise degenerate states.
However, this does not affect the 
branching ratios for the flavour-violating radiative decays.
Therefore, we simply use~\cite{gllv}
\begin{equation}
\tilde{m}_{\nu}^2= (m_{\tilde{l}}^\delta)^2+ \delta m_{\nu_D}^2 + \frac{1}{2}
M_Z^2 cos 2\beta
\end{equation}
in our subsequent calculations.

Next we consider the rare decay $\mu \ra 3e$ and $\mu \ra
e$ conversion on nuclei, starting with the $\mu \ra 3e$ reaction.
This reaction is interesting on its own since it has a structure
 much richer than that of the $\mu\ra e\gamma$ decay. Thus,
it is possible that  $\mu\ra 3 e$ can take place in cases where
$\mu\ra e\gamma$ is forbidden.
This and the related $\tau \ra 3 \mu/e$ decay receive contributions from
three types of Feynman diagrams.
The first are photon `penguin' diagrams related to
the diagrams for
$\mu \ra e \gamma$ and $\tau \ra \mu/e \gamma$ discussed above, where now
the photon is virtual and decays into an $e^+e^-$ (or $\mu^+ \mu^-$) pair.
A second class of diagrams is obtained
by replacing the photon line with a $Z$ boson. Finally, there are also
box diagrams. In addition, all the above types of diagrams are accompanied
by their supersymmetric analogues. We evaluate
all the relevant diagrams
exactly in our subsequent numerical
analysis. However,  we note that
the dominant diagrams in the models of interest to us here are generally
the `penguin' diagrams with an intermediate
off-shell photon, which contribute via the $A_M$ and $A_E$ terms
presented in (\ref{AM},\ref{AE}).
Compared to $\mu\ra e \gamma$ decay, the branching ratio is,

\begin{equation}
{\Gamma ( \mu^+ \ra e^+e^+e^-) \over \Gamma ( \mu^+ \ra e^+ \gamma)}
\approx 6 \times 10^{-3}
\label{approxratio}
\end{equation}
This does not necessarily mean that
$\mu \ra 3 e$ decay is uninteresting to experiment, because the
experimental detection and background problems are very different for the
two decays. However, we do not present
detailed numerical results for $\mu^+ \ra e^+e^+e^-$ decay, because the
factor (\ref{approxratio}) is essentially universal. There is a similar
small ratio for $\tau \ra 3 \mu/e$ relative to $\tau \ra \mu/e \gamma$,
which seems to preclude its observation even at the LHC.

The $\mu \ra e$   conversion is a coherent process in a muonic atom
originally studied in~\cite{FW,CG}.
Even though this reaction proceeds via
the same classes of diagrams
as those discussed above in connection with $ \mu^+ \ra e^+e^+e^-$,
this reaction is rather different, since it involves hadronic currents.
Effects of nuclear nature, such as  the size of the nucleus  in particular when
heavy atoms are involved, play important role. 
Detailed calculations including all contributions from penguin and
 the box diagrams will be given in section 5.  Here,
in order to obtain a first rough estimate 
of the ratio of the $\mu \ra e$ conversion to the $\mu\ra e\gamma$ reaction,
we restrict to the photonic contribution which dominates over a large
portion of the parameter space.  The  $\mu \ra e$ conversion rate relative to
conventional muon capture, 
 is given by
\begin{eqnarray}
R(\mu \ra e) \equiv 
\frac{\Gamma(\mu^+\ra e^+)}{\Gamma(\mu\ra \nu_{\mu})}&=&
(\frac{4\pi\alpha}{G_F})^2 \frac{E_ep_e}{m_{\mu}^2}
\frac{|ME|^2}{C Z f(A,Z)} (|A^L+A^R|^2+|A^L-A^R|^2)
\label{conversion}
\end{eqnarray}
where $A^{L/R} \equiv A^{L/R}_M+A^{L/R}_E$. It is important to
note that the combination of matrix elements $A_{M/E}^{L/R}$
in (\ref{conversion}) is different from that in appearing
$\mu \ra e$ decay (\ref{BR}). The function
$f(A,Z)$ is given in~\cite{prima},
and has the following approximate value for
elements with $A\approx 2 Z$  
\bea
f(A,Z)= 1.0-0.03 \frac{A}{2Z}+0.25 \left(\frac{A}{2 Z}-1\right)+3.24
\left(\frac{Z}{2A}-\frac 12-\frac{1}{4 A}|\frac{A}{2Z}-1|\right)
\approx 0.16
\label{primakov}
\eea
Further, the parameter $C$ in (\ref{conversion})
is~\cite{prima}
\bea
C=|F_V^2+3 F_A^2 + F_P^2 -2 F_A F_P|\sim 5.9
\eea
and the nuclear matrix element  $|ME|$ is
\begin{equation}
|ME|=|<(A,Z)_{f}|J(0)|(A,Z)_i>|\approx Z F_{c}(q^2),
\label{me}
\end{equation}
where, in the case of the photonic diagram,
$ F_{c}(q^2)$ stands for the proton or neutron form factor. Calculations
using the nuclear shell model for ${}^{48}{}_{22}Ti$
have yielded
$F_{c}(q^2=-m_{\mu}^2)\approx 0.543$ for the case of proton and $0.528$ for the
neutron~\cite{Kosmas:1997eb,Kosmas:1997ec}. Comparing now with
$B(\mu^+\ra e^+\gamma)$,
one obtains a rough estimate of the expected range of the $\mu \ra e$
conversion ratio:
\begin{eqnarray}
 R({\mu^+ Ti\ra e^+ Ti})&\approx&
\frac{\alpha}{3\pi}\frac{E_ep_e}{m_{\mu}^2}
\frac{Z F_c^2}{C  f(A,Z)}BR(\mu\ra e\gamma)\nonumber\\
  &\approx & 5.6 \times 10^{-3} BR(\mu\ra e\gamma)
\label{convratio}
\end{eqnarray}
which shows a relative suppression about two orders of magnitude.
Nevertheless, $\mu \ra e$ conversion is also  interesting.
Current experimental bounds give 
$R(\mu^+ Ti \ra e^+ Ti)\le 6.1\times 10^{-13}$,
whilst ongoing experiments will reach $\sim 2\times 10^{-14}$.
However, with an intense proton (and muon) source, such as that
projected for a neutrino factory or a muon collider,
experiments sensitive to rates as low as $10^{-16}$
may be feasible~\cite{mufact}. Moreover, as is apparent from
(\ref{conversion}) above, and the appearance of box diagrams,
the structure of the matrix element is
different from that for $\mu \ra e \gamma$ (\ref{BR}), so the ratio
(\ref{convratio}) is not universal, unlike the $\mu \ra 3 e$
case. Therefore, although we emphasize $\mu \ra e \gamma$
in what follows, we shall also present later some results for
$\mu \ra e$ conversion on Titanium.

\section{Neutrino Mass Textures in the Light of Super-Kamiokande}

Having discussed the theoretical framework for calculating
charged-lepton-flavour violation in a supersymmetric model
with universal input soft supersymmetry-breaking parameters
at the GUT scale, we now discuss possible extreme patterns of
neutrino masses, mixings and Yukawa couplings that might figure in
the renormalization-group equations.

The most basic piece of information from
the recent atmospheric-neutrino data is the existence
of at least one large mixing angle in the
lepton sector, that associated with the
$\mu-\tau$ flavour mixing.
Lepton mixing may in general
arise either from the charged-lepton sector,
or the neutrino sector, or both. 
In analogy to the quark mixing
matrix $V_{CKM}$, the leptonic mixing matrix
$V_{MNS}$ is defined as~\cite{mns}
\bea
V_{MNS} =
V_{\ell}V_{\nu}^{\dagger}
\label{vmns}
\eea
where $V_{\ell}$ 
transforms the left-handed charged leptons
to a diagonal mass basis, 
whereas
$V_{\nu}$ diagonalizes the light-neutrino mass matrix $m_{eff}$.
In the see-saw framework 
which provides a natural mechanism for generating
very light neutrinos \cite{seesaw}, the latter
is given by
\begin{equation}
m_{eff}
=m^D_{\nu}\cdot (M_{\nu_R})^{-1}\cdot m^{D^{\normalsize T}}_{\nu}.
\label{eq:meff2}
\end{equation}
where $m^D_{\nu}$ and 
$M_{\nu_R}$ stand for the Dirac and
the heavy Majorana neutrino mass matrices
respectively.

How may one characterize the structures of Dirac and 
heavy Majorana matrices that generate  viable 
neutrino textures consistent with the
Super-Kamiokande data? In \cite{ELLN2},
we proposed a classification according to the criteria
of `matched' and `mismatched mixing', as defined below.

$\bullet$ {\em Matched mixing}:
This occurs when there is only one large 
neutrino mixing angle, namely that
in the (2--3) sector of the light
neutrino mass matrix $m_{eff}$,
as suggested by the atmospheric neutrino data,
and there is no large mixing
in other sectors of either
the light Majorana or the Dirac neutrino 
mass matrix. In this case, the
problem reduces approximately to a $2\times 2$ mixing problem.
Since there is no way to render
three degenerate neutrinos consistent with
the bounds from neutrinoless double beta decay
without also large (1-2) neutrino mixing,
`matched mixing' requires hierarchical 
neutrino masses. 

In this case, it has been
shown that (in the absence of zero-determinant
solutions, i.e., solutions where 
strong cancellations in the (2,3) sub-determinant in the
neutrino sector cause one of the eigenvalues
to be relatively small, which we will discuss
in an example below) the lepton mixing originates entirely from
the Dirac mass matrices,
while the structure of the heavy Majorana mass
matrix $M_{\nu_R}$ does not affect the low-energy lepton mixing 
\cite{GGR2}. This result remains valid even when
renormalisation group effects are taken into account
\cite{CELW}. Then, in the basis where the
charged lepton flavours are diagonal, the Dirac neutrino 
mass matrix takes a very simple form, given by
\bea
m^D_\nu \propto  \left
(\begin{array}{ccc}
0 & 0  & 0 \\
0 & x^2 & x \\
0 & x & 1
\end{array}
\right), ~~
m^D_\nu \propto  \left
(\begin{array}{ccc}
0 & 0  & 0 \\
0 & xy & x \\
0 & y & 1
\end{array}
\right)
\eea
for symmetric  and asymmetric textures, respectively~\footnote{Here
we have correlated the (2-2) with the 
(2-3) and (3-2) elements, assuming that they arise from
a flavour symmetry of the type discussed later.}.

It is evident that, in such a scenario,  the (1-2) mixing and
(1-3) mixing are both
zero, in first approximation,
so the large (2-3) mixing is not
communicated at all to the (1-2) sector,
for which even the sub-dominant contributions
are very small. Passing to the (2-3) mixing,
we see that it can be either maximal or
non-maximal, depending on
the value of $x$, which can be as large as
unity. It is clear that such a `matched-mixing' scenario is
consistent only with the small-mixing-angle (SMA) MSW
solution for solar neutrinos, since the large-mixing-angle (LMA)
MSW solution and the vacuum-oscillation (VO) scenarios both
require large (1--2) and/or (1--3) mixing.

If (1--2) and (1--3) mixing are both small, as in
a generic `matched-mixing' model, the $\mu \rightarrow e \gamma$ and
$\tau \rightarrow e \gamma$ rates
should be relatively `small', whereas the $\tau \rightarrow \mu \gamma$
rate may be relatively `large'. These general expectations are
borne out in the model calculations presented later.

$\bullet$ {\em Mismatched mixing}:
Entirely different structures arise when 
(i) there is more than one large
mixing angle in $m_{eff}$, and/or
(ii) there is a large
Dirac mixing angle that involves
different generations from those of the light Majorana matrix.
A mild example of mismatched mixing occurs, for example, when the
atmospheric problem is solved
by $\nu_{\mu}\ra \nu_{\tau}$ oscillations, whilst the Dirac mass
matrix is related to the quark mass matrix, with Cabibbo-size mixing
between the first and second
generations. 

In generic mismatched mixing models,
there are relatively large violations of charged-lepton
flavour in all the (1--2), (2--3) and (3--1) channels.
Thus, such a `mismatched-mixing' scenario is {\it a priori}
compatible with either the LMA or VO solutions of the
solar-neutrino problem. Moreover,
$\mu \ra e \gamma$ and $\tau \ra e \gamma$
generically have larger rates than they would have in matched-mixing
models. However,
the structure of the Majorana matrix
becomes more complicated in the case of mismatched mixing.
In particular, it
is possible that the Dirac mass matrix is almost diagonal,
with a large hierarchy of Dirac couplings,
so that, in particular, 
$\lambda_1 \ll \lambda_2$ where the $\lambda_i$ are the
eigenvalues of the neutrino Dirac coupling matrix.
In this case, 
the light entry of the heavy Majorana mass matrix again
effectively decouples from the heavier ones~\cite{ELLN2}.
However, this is no longer true if
the (1--2) mixing angle in the
Dirac mass matrix increases.

One particularly interesting example
of such large mixing, which has been extensively studied in
the literature, is that of bi-maximal mixing~\cite{bimaximal}, for which
one can easily obtain viable solutions with degenerate neutrinos.
As an example, we quote the texture 
\bea
{m_{eff}} \propto \,\pmatrix{
0 &  {1\over\sqrt2} &  {1\over\sqrt2}\cr
{1\over\sqrt2} &  {1\over2} &  -{1\over2}\cr
{1\over\sqrt2} &  -{1\over2} &  {1\over2}\cr
}
\label{GGtexture}
\eea
For this texture, in the neutrino
mixing parametrization
\bea
\pmatrix{\nu_e \cr \nu_\mu\cr \nu_\tau\cr}=
\pmatrix{c_2c_3 &   c_2s_3 &   s_2e^{-i\delta}\cr
-c_1s_3-s_1s_2c_3e^{i\delta} &  
+c_1c_3-s_1s_2s_3e^{i\delta} &  s_1c_2\cr
+s_1s_3-c_1s_2c_3e^{i\delta} &  
-s_1c_3-c_1s_2s_3e^{i\delta} &   c_1c_2\cr}\,
\pmatrix{\nu_1\cr \nu_2\cr \nu_3\cr}\,,
\label{param}
\eea
where the diagonal matrix $m_{eff}^{diag}$ is 
$diag(m_1e^{i\phi}, m_2e^{i\phi'}, m_3)$, and
$\phi$ and
$\phi'$ are phases in the light Majorana mass matrix,
one finds that the mixing angles can be
\bea
\phi_1 = {\pi \over 4},~~
\phi_2 = 0, ~~\phi_3 = {\pi \over 4}
\eea
However, this mixing is not stable
under perturbations of the degenerate texture (\ref{GGtexture})~\cite{EL}.
After including renormalisation group
effects, the associated mixing angles
become~\footnote{Renormalisation-group effects on
schemes with neutrino degeneracy have also
been discussed in~\cite{EC}.}
\bea
\phi_1 \approx -0.327, ~~\phi_2 \approx 0.415, ~~
\phi_3 \approx  -0.884
\eea
which is inconsistent with a degenerate mass scale
much above 1~eV. However, ways around this difficulty have
been proposed in the context of non-Abelian flavour symmetries,
as discussed later.

\section{Neutrino Masses from Flavour Symmetries}

\subsection{Examples based on Abelian Groups}

In this section we give a brief description of models
based on extra Abelian symmetries which lead to a 
consistent charged fermion mass spectrum and give
predictions for the neutrinos and flavour violating processes. 
The fact that the fermion mass matrices exhibit a hierarchical
structure suggests that they are generated by 
an underlying family symmetry, of which
the simplest examples are based on Abelian 
groups. To review how the various terms in the mass matrices arise in
such a model, we first denote
the  charges of the Standard Model 
fields under the symmetry as in Table 1.
The Higgs charges are chosen so that the
terms $f_3 f^c_3 H$ (where $f$ denotes a fermion
and $H$ denotes
$H_{1}$ or $H_{2}$) have zero charge.
Thus, when the $U(1)$ symmetry is unbroken only the (3,3) elements
of the associated mass matrices will be non-zero. 
When the $U(1)$ symmetry is spontaneously broken via 
standard model singlet fields,
$\theta,\; \bar{\theta}$, with opposite $U(1)$ charge
and equal vevs (vacuum expectation values),
the remaining entries are generated in a hierarchical manner.
The suppression factor for each entry depends on the family charge:
the higher the net $U(1)$ charge of a term $f_i f^c_j H$,
the higher the power $n$ in a non-renormalizable term
$f_i f^c_j H \left ( \frac{\theta}{M} \right)^n $
that has zero charge. Here $M$ is a mass 
scale associated with the mechanism that generates 
the non-renormalizable terms. A common approach
communicates symmetry breaking via an 
extension of the `see-saw' mechanism,
mixing light and heavy states,  known as the 
Froggatt--Nielsen mechanism~\cite{FN}. 

\begin{table}[h]
\begin{center}
\begin{tabular}{|c|cccccc|}
\hline
& $Q_{i}$ & $u_{i}^{c}$ & $d_{i}^{c}$ & $L_{i}$ & $e_{i}^{c}$ & $\nu
_{i}^{c} $ \\ \hline
$U(1)$ & $\alpha _{i}$ & $\beta _{i}$ & $\gamma _{i}$ & $b_{i}$ & $%
c_{i} $ & $d_{i}$ \\ \hline
\end{tabular}
\end{center}
\caption {\it Notation for the $U(1)$ charges of the Standard Model
fields, where $i$ stands for a generation index.}
\end{table}

We discuss the simplest possible scheme, with
symmetric mass matrices~\cite{IR}~\footnote{The 
lepton sector in this case is
identical to that of
$SU(3)_{c}\times SU(3)_{L}\times SU(3)_{R}$~\cite{GGR2}, which, however,
predicts asymmetric quark
mass matrices.}. This leads to three viable cases with
charges \cite{GGR2}
\begin{eqnarray}
A):&b_{i}=c_{i}=d_{i} &= \left (-\frac{7}{2},\frac{1}{2},0 \right ) 
\nonumber  \\
B):&b_{i}=c_{i}=d_{i} &= \left (\frac{5}{2},\frac{1}{2},0 \right ) 
\nonumber   \\
C):&b_{i}=c_{i}=d_{i} &=(3,0,0)
\label{char}
\end{eqnarray}
leading to three possible charged-lepton matrices :
\bea
M_{\ell }  \propto  \left( 
\begin{array}{ccc}
\bar{\epsilon}^{7} & \bar{\epsilon}^{3} & \bar{\epsilon}^{7/2} \\ 
\bar{\epsilon}^{3} & \bar{\epsilon} & \bar{\epsilon}^{1/2} \\ 
\bar{\epsilon}^{7/2} & \bar{\epsilon}^{1/2} & 1
\end{array}
\right) ,
M_{\ell } &  \propto & \left( 
\begin{array}{ccc}
\bar{\epsilon}^{5} & \bar{\epsilon}^{3} & \bar{\epsilon}^{5/2} \\ 
\bar{\epsilon}^{3} & \bar{\epsilon} & \bar{\epsilon}^{1/2} \\ 
\bar{\epsilon}^{5/2} & \bar{\epsilon}^{1/2} & 1
\end{array}
\right)  \nonumber \\
M_{\ell } & \propto & \left( 
\begin{array}{ccc}
\bar{\epsilon}^{6} & \bar{\epsilon}^{3} & \bar{\epsilon}^{3} \\ 
\bar{\epsilon}^{3} & 1 & 1 \\ 
\bar{\epsilon}^{3} & 1 & 1
\end{array}
\right) 
\label{LEPTONS} 
\eea
The first two matrices lead to
natural lepton hierarchies for $\bar{\epsilon} \approx 0.2$ 
and imply large but non-maximal lepton mixing.
On the other hand,
the third matrix leads to maximal mixing in the (2-3) sector, but
requires an
accurate cancellation in order to get the correct
ratio $m_{\mu }/m_{\tau }$.

The neutrino Dirac mass is specified to have
the same form as the charged leptons, but with a different
expansion parameter. Indeed,
since neutrinos and up-type quarks
(charged leptons and down-type quarks)
couple to the same Higgs, they should have the same expansion 
parameter $\epsilon (\bar{\epsilon})$,
where the spread between the up- and down-quark hierarchies
requires $\epsilon \approx  \bar{\epsilon}^2 \approx 0.05 $~\cite{IR}.
Then, 
\bea
m^D_{\nu} \propto \left( 
\begin{array}{ccc}
{\epsilon}^{7} & {\epsilon}^{3} & {\epsilon}^{7/2} \\ 
{\epsilon}^{3} & {\epsilon} & {\epsilon}^{1/2} \\ 
{\epsilon}^{7/2} & {\epsilon}^{1/2} & 1
\end{array}
\right),~
m_{\nu }^D \propto \left( 
\begin{array}{ccc}
{\epsilon}^{5} & {\epsilon}^{3} & {\epsilon}^{5/2} \\ 
{\epsilon}^{3} & {\epsilon} & {\epsilon}^{1/2} \\ 
{\epsilon}^{5/2} & {\epsilon}^{1/2} & 1
\end{array}
\right)  \label{Diracmasses}
\eea
for the first two choices of charges in 
(\ref{char}) respectively.

The mass structure of the light neutrinos is more
complicated, due to the heavy Majorana masses of the
right-handed components. However, as discussed above,
the neutrino mixing in the case of large neutrino
hierarchies and non-zero determinant solutions
is determined entirely by the 
Dirac mass matrices (which feel the left-handed 
charges). The heavy Majorana mass sector only
affects the neutrino eigenvalues.
For instance, for the set B of solutions, with
$b_{i}=c_{i}=d_{i} = \left (\frac{5}{2},\frac{1}{2},0 \right ) $,
we find that~\cite{GGR2,M3}
\bea
V_\ell = \left( 
\begin{array}{ccc}
1 & \bar{\epsilon}^{2} & -\bar{\epsilon}^{5/2} \\ 
-\bar{\epsilon}^{2} & 1 & \bar{\epsilon}^{1/2} \\ 
\bar{\epsilon}^{5/2} & -\bar{\epsilon}^{1/2} & 1
\end{array}
\right), V_{\nu_D} = \left( 
\begin{array}{ccc}
1 & \bar{\epsilon}^{4} & -\bar{\epsilon}^{5} \\ 
-\bar{\epsilon}^{4} & 1 & \bar{\epsilon} \\ 
\bar{\epsilon}^{5} & -\bar{\epsilon} & 1
\end{array}
\right) \label{Bsolutions} 
\eea
whilst for the set A of solutions with
$b_{i}=c_{i}=d_{i} = \left (-\frac{7}{2},\frac{1}{2},0 \right)$
we find that
\bea
V_\ell = \left(
\begin{array}{ccc}
1 & \bar{\epsilon}^{2} & -\bar{\epsilon}^{7/2} \\
-\bar{\epsilon}^{2} & 1 & \bar{\epsilon}^{1/2} \\
\bar{\epsilon}^{7/2} & -\bar{\epsilon}^{1/2} & 1
\end{array}
\right), V_{\nu_D} = \left(
\begin{array}{ccc}
1 & \bar{\epsilon}^{4} & -\bar{\epsilon}^{7} \\
-\bar{\epsilon}^{4} & 1 & \bar{\epsilon} \\
\bar{\epsilon}^{7} & -\bar{\epsilon} & 1
\end{array}
\right) \label{Asolutions}
\eea
We see that both of these solutions are close to the
limit of `matched' mixing discussed in the previous section,
since the mixing in the (1-2) and (1-3) sectors is much
smaller than in the (2-3) sector, even though the latter may
be less than maximal. 

The solution C is also close
to the limit of `matched' mixing, but with maximal (2-3)
mixing. The lepton mixing matrix in this case is
\bea
V_\ell \approx \left(
\begin{array}{ccc}
\frac{1}{\sqrt 2} & -\frac{1}{\sqrt 2}&\bar{\epsilon}^{3}\\
-\frac{1}{2}&-\frac{1}{2}&-\frac{1}{\sqrt 2}\\
\frac{1}{2}&\frac{1}{2}&-\frac{1}{\sqrt 2}
\end{array}
\right)
\label{thirdVl}
\eea 
and $V_{\nu_D}$ has a similar form, with the small entry
${\bar \epsilon}^3 \ra {\bar \epsilon}^6$, in analogy with
(\ref{Bsolutions}, \ref{Asolutions}).
The diagonalization of the lepton mass matrix leads in this case to
eigenvalues $\bar{\epsilon}^{3}, -\bar{\epsilon}^{3},2$, and some
fine-tuning would be needed to obtain correct low-energy masses.
In this case, $V_{\ell}$ might differ from (\ref{thirdVl}).

\subsection{Examples based on non-Abelian Groups}

We do not discuss these type of models in detail, but
remark that they lead naturally to
models with degenerate neutrinos.
Indeed, when the three lepton
doublets form a real irreducible representation of some non-Abelian
flavour group, as for instance a triplet of $SO(3)$,
which has been extensively studied during the last year~\cite{nonabel},
one expects exact neutrino mass degeneracy at zeroth order.
The same result can be achieved by discrete non-Abelian
symmetries~\cite{discrNA}
and  is to be contrasted to the predictions
of abelian flavour symmetries,
which naturally lead to large hierarchies between
the various mass entries, as reviewed above. However,
the non-Abelian symmetry is typically broken by terms
related to lepton masses.
Once the flavour symmetry is broken, the exact neutrino
mass degeneracy is lifted by small terms and 
one may be able to reproduce the Super-Kamiokande
data.

Generic neutrino mass textures that originate
from such flavour symmetries are of
the `mismatched-mixing' type,
as in the example (\ref{GGtexture}) mentioned above.
It was pointed out~\cite{EL} that such textures are vulnerable
to radiative corrections, which may lead to
unacceptable patterns of masses and mixing angles.
The requirement
that such a disaster be avoided
imposes severe  constraints
on the mixing angles and requires the
mixing should be close to bi-maximal~\cite{BRSt}.
Thus, one would expect a large mixing angle in
the $\mu-e$ flavour sector, which would
tend to generate relatively large
large rates for $\mu \rightarrow e \gamma$,
$\mu \rightarrow 3 e$ and $\mu \ra e$ conversion,
as well as for $\tau \ra \mu \gamma$ decay.

\subsection{Flipped-$SU(5)$ Model derived from String}

We now outline how the above 
analysis may be extended to a typical
grand-unified model derived from string. In such a framework, 
the following features appear generic:
(i) non-Abelian symmetries are disfavoured,
(ii) the $U(1)$ symmetries and charges are
specified in any given string model, and one generically
expects a product of anomalous and non-anomalous Abelian groups, rather
than a single $U(1)$,
(iii) there are many singlet fields
involved in the mass generation,
not just a single pair $\theta$, $\bar{\theta}$. 
Once a string model is chosen, e.g., by specifying fermionic
boundary conditions on the world sheet, then automatically
the gauge properties of the model and the 
quantum numbers of all fields, including those
which may acquire non-zero vevs and fill in the
fermion mass matrices, are specified.
The field vevs that determine the
magnitudes of the various entries
are constrained by the anomaly-cancellation conditions 
and the flat
directions of the effective potential in the theory.
Finally, we recall that
(iv) additional string symmetries
(expressed through calculational selection
rules~\cite{select}) further constrain
the possible forms of the mass matrices, since they
forbid most of the Yukawa couplings that are 
allowed by the rest of the symmetries of
the model.

An example of the above class of models is provided by the 
flipped $SU(5)$ model derived from string~\cite{aehn}, 
specializing to the pattern of vevs and mass matrices
discussed in~\cite{ELLN,ELLN2}.
Looking at the field assignments in group representations,
one sees that:
(i) since the charge conjugates of the right-handed neutrinos have the
same charges as the down quarks, the Majorana mass matrix will be
constrained, and (ii) due again to the common charge
assignments, the Dirac neutrino mass matrix is the transpose of the up-quark
mass matrix.
The quark and charged-lepton mass matrices have been presented
in~\cite{ELLN}, where 
the possible flat directions of the theory
were also reconsidered. 
Since the analysis of the surviving couplings
after all symmetries and string selection rules
are taken into account is quite involved,
we refer to the original references for
details, while here we just give
an illustration of the predictions for neutrino masses.

Within this model, we found that the charged-lepton mixing matrix 
is given by
\beq
V_{\ell} = \left (
\begin{array}{ccc}
1-\frac{1}{2} x^2 & x & 0 \\
-x &  1-\frac{1}{2} x^2 & 0 \\
0 & 0 & 1
\end{array}
\right )
\label{veenum}
\eeq
where $x$ is the vev of a combination of 
hidden-sector fields that transform as
sextets under $SO(6)$, that needs to be
${\cal O}(1)$ for realistic quark mass
matrices~\cite{ELLN}.
The Dirac neutrino mass matrix, $m_{\nu}^D$,  is expressed
in terms of three expansion parameters $f,x,y$ related to field vevs
scaled by the string mass:
\begin{equation} 
m_{\nu}^D = 
\left (
\begin{array}{ccc}
x f & 1 & 0 \\
f & x & 0 \\
0 & 0 & y
\end{array}
\right) 
\label{Dira}
\end{equation}
where $y$ stands for the $SU(5)$ decuplets that break
the gauge group down to
the Standard Model, with scaled  vev
$\approx M_{GUT}/M_{s}$, where
$M_s$ is the string scale. In weakly-coupled string constructions,
this ratio is suppressed, but
the GUT and the string scales can coincide
in the strong-coupling limit of
$M$ theory, in which case $y \approx 1$. Finally,
$f$ stands for a singlet field, the value of whose
vev $\approx 0.04$ is fixed in order of magnitude by
the quark mass  hierarchies.

Since now the mixing in the (2-3) sector of the
Dirac mass matrix is zero, we see that in
order to generate a large hierarchy among
the various neutrinos, we necessarily
require zero-determinant solutions.
In~\cite{ELLN2}, where the expectations
for neutrino masses were studied, we
ended up with two possible forms
for $M_{\nu_R}$, depending
on the vevs of the singlet fields. These were
\bea
M_{\nu_R} \propto
\left (
\begin{array}{ccc}
\alpha & 0 & 0 \\
0 & 0 & f y \\
0 & f y & t x
\end{array}
\right ), ~~
M_{\nu_R} \propto 
\left (
\begin{array}{ccc}
f y^2  & \lambda x y^2  & 0 \\
\lambda xy^2  & 0 & f  y \\
0 & f y & t x
\end{array}
\right) \label{maj2}
\eea
where in the second example a notional factor of $\lambda\approx {\cal O}(1)
$ has been included
so as to avoid sub-determinant cancellations,
which are not expected to arise once 
coefficients of order unity are properly taken into account.

As can be seen from the lower $2\times 2$ matrices in (\ref{maj2}),
consistency with the neutrino data implies
$y \approx 1 $, as could occur in the strong-coupling limit 
of $M$ theory, and  $ t \sim f$. The parameter $\alpha$ 
represents a higher-order non-renormalizable contribution
$\left(\langle\Phi\rangle \over M_s\right)^n$, where $\Phi$ is an  
effective singlet, that is expected to appear at some power $n>7$. Its
actual value is irrelevant, provided that it not too tiny,
in which case $m_{\nu_1}$ might be increased to
an unacceptable value. 

The situation described just above corresponds to the `mismatched'
scenario introduced above and exemplified previously in the context of
simpler $U(1)$ models.
It is straightforward now to determine the effective light neutrino mass
matrix
for the above cases, and see that large mixing is implied for the 
$\nu_{\mu}-\nu_{\tau}$ sector, as required by the atmospheric neutrino
data. We do not
go into further details, since they would depend on more specific
aspects of the model, which contains several poorly-constrained
expansion parameters, not all of which are necessarily very small.
However, we can infer some qualitative properties of the predictions of
this flipped $SU(5)$ model for flavour-violating decays. From 
the form of the charged-lepton mass matrix  (\ref{veenum}), we 
would expect
a rather large $\mu\ra e\gamma$ branching ratio. On the 
other hand, the $\tau$ lepton remains completely decoupled from the
lighter families in the approximation considered here~\footnote{This 
is, of course, connected to the fact that the large
$\nu_{\mu}-\nu_{\tau}$ mixing needed to interpret the atmospheric
neutrino data comes entirely from the effective light Majorana matrix.}.
This is in contrast with many of the $U(1)$ models, with the
result this flipped $SU(5)$ model would predict
a relatively small branching ratio for $\tau\ra \mu\gamma$.

\section{Predictions for Rare Processes}

We now discuss quantitatively implications of the large neutrino mixing
needed to interpret the neutrino data
for processes that violate the
conservation of charged lepton flavours.
In particular, as we shall see, the 
likelihood that the atmospheric neutrino problem is solved by
$\nu_{\mu}-\nu_{\tau}$ mixing suggests the likely appearance of the
$\tau\ra\mu \gamma$ transition at a rate that may be
accessible~\cite{LHC}.
This observation supplements the common
belief that $\mu\ra e \gamma$ and related processes may offer
good prospects for observing charged-lepton-flavour violation.

In this Section we present results 
for these and related
flavour-changing decays for generic examples of the $U(1)$
textures discussed in the previous Sections.
We start with the radiative decays $\mu \ra e \gamma$ and $\tau \ra \mu
\gamma$ \cite{neutrinoLFVs}.
For the reasons discussed earlier, we do not explore further
the corresponding decays where the photon is replaced by
an $e^+ e^-$ pair.
However, we do present later some numerical results for $\mu-e$
conversion on $Ti$, whose rate is not directly related to that for
$\mu \ra e \gamma$, and which may present some experimental
advantages \cite{KO}.

In addition to the flavour-mixing effects, the rates for flavour
non--conserving decays are also
sensitive  to other physical  quantities. In particular, in the
supersymmetric GUT context explored here, they depend on the masses 
of the sparticles that
mediate   the flavour non-conserving processes.
We parametrize their masses in terms of the universal GUT-scale
parameters $m_0$  and $m_{1/2}$, and use the renormalization-group
equations of the MSSM to calculate the low-energy sparticle
masses, taking into
account low-energy threshold effects.  Other relevant free parameters of
the
MSSM (discussed in detail in~\cite{gllv}) are the trilinear coupling $A$,
the sign of the Higgs mixing parameter $\mu$, and
the value of $\tan\beta$.  Here we fix the value of  $A_0=- m_{1/2}$, 
assume the
sign of the $\mu$ parameter to be either positive or negative, and restrict our
analysis
to low and intermediate values of $\tan\beta \le 10$. 
Before presenting our results, we first consider in more detail
the relevance of some of the parameters entering in the
calculation.

$\bullet$
The case of non-universal soft
masses at the GUT scale was analysed in~\cite{gllv}. However,
in such a case, the predictions
for $\mu\ra e \gamma$ exceed the current limits for most of the
parameter space. this is why, in the present work, we restrict ourselves
to the case
of universal soft masses at the GUT scale,  assuming
the existence of some mechanism that
assures universality and the absence of mixing in the K\"ahler
potential.

$\bullet$
No constraints on the $m_0, m_{1/2}$ values were
discussed in~\cite{gllv}. Here, however, we choose their initial values
so as to
respect the cosmological relic-density constraints, as discussed in
\cite{ES}.

$\bullet$
To calculate within a given fermion mass texture,
we need to estimate the mixing both of the neutrinos and of
the charged leptons.
In the case of $U(1)$ models, all the mass matrices are given as
power expansions in the parameters $\epsilon, \overline{\epsilon}$, and
it is not possible to fix uniquely the ${\cal O}(1)$ coefficients $c_{ij}$ needed
to obtain masses and mixings.  
There is a similar ambiguity in flipped $SU(5)$ models.

For our numerical study here, we adopt for  the three
$U(1)$ models discussed in Section 4 an approach 
similar to that of~\cite{gllv}, commenting later on the
expectations for other classes of models. In~\cite{gllv}, a
 set of ${\cal O}(1)$ coefficients $c_{ij}$ was used which led to the
correct mass spectrum, with the  mixing in the charged-lepton 
mass matrix emerging as a prediction. 
Nevertheless,  the choice of $c_{ij}$ is not unique, since many
sets
can give the same mass eigenvalues but different mixing. In order to test
the sensitivity of our results to this arbitrariness, here
we use three representative sample sets of coefficients
$c_{ij}\equiv c_{ji}$  for each of the $U(1)$ textures presented in Section 4, 
namely
\begin{eqnarray}
\label{cij}
A)\,\, c_{12}&=&.87 ,\,c_{22}=.59 ;\;\;
  c_{12}=.89,\,c_{22}=1.44 ;\;\;
 c_{12}=.61, c_{23}=.79\nonumber\\
B)\,\, c_{12}&=&1.68 ,c_{22}=.56 ;
  \;\;c_{12}=1.70,\,c_{22}=1.45;\;\;
 c_{12}=1.41, c_{23}=.79
\nonumber\\
C)\,\, c_{12}&=&2.59 ,c_{22}=1.27
  ;\;\; c_{12}=2.34,\,c_{22}=.79\;\;
 c_{12}=2.75, c_{23}=1.13
\nonumber
\label{choices}
\end{eqnarray}
which we denote hereafter by $A_{1,2,3}, B_{1,2,3}$ and $C_{1,2,3}$,
respectively.

We plot in Figs.~2, 3 and 4 the branching ratios for
$\mu\ra e\gamma$ and $\tau\ra \mu \gamma$
in the textures A, B and C, respectively,
assuming in each case the representative value
$m_{1/2}=250$ GeV, $\tan\beta =3$ and allowing both possible signs of
$\mu$. In each of Figs.~2, 3 and 4 the spread in the numerical results
reflects the uncertainties associated with the 
numerical coefficients $c_{ij}$. This spread immediately warns us
that the results shown in
Figs.~2, 3 and 4 should be understood only as
order-of-magnitude estimates. We note
that the predictions for the
$\mu\ra e\gamma$ decay branching ratio in textures A and B
are particularly sensitive to
these coefficients, whilst $\tau\ra \mu \gamma$ reaction is 
generally less sensitive, as also is $\mu \ra e \gamma$
in texture C.
In all cases, the $\mu >0$ choices (solid curves) lead to an
enhancement of  the branching ratios compared to
the choice $\mu <0$, at least for larger values of $m_0$.  We note also
that,  for a small range of $m_{0}$ values that depends on the 
 texture and the specific choice of the coefficients,
there is potentially a strong cancellation of the mixing effects,
leading to a considerable suppression of the branching ratio.

\begin{figure}
\begin{minipage}[b]{8in}
\epsfig{file=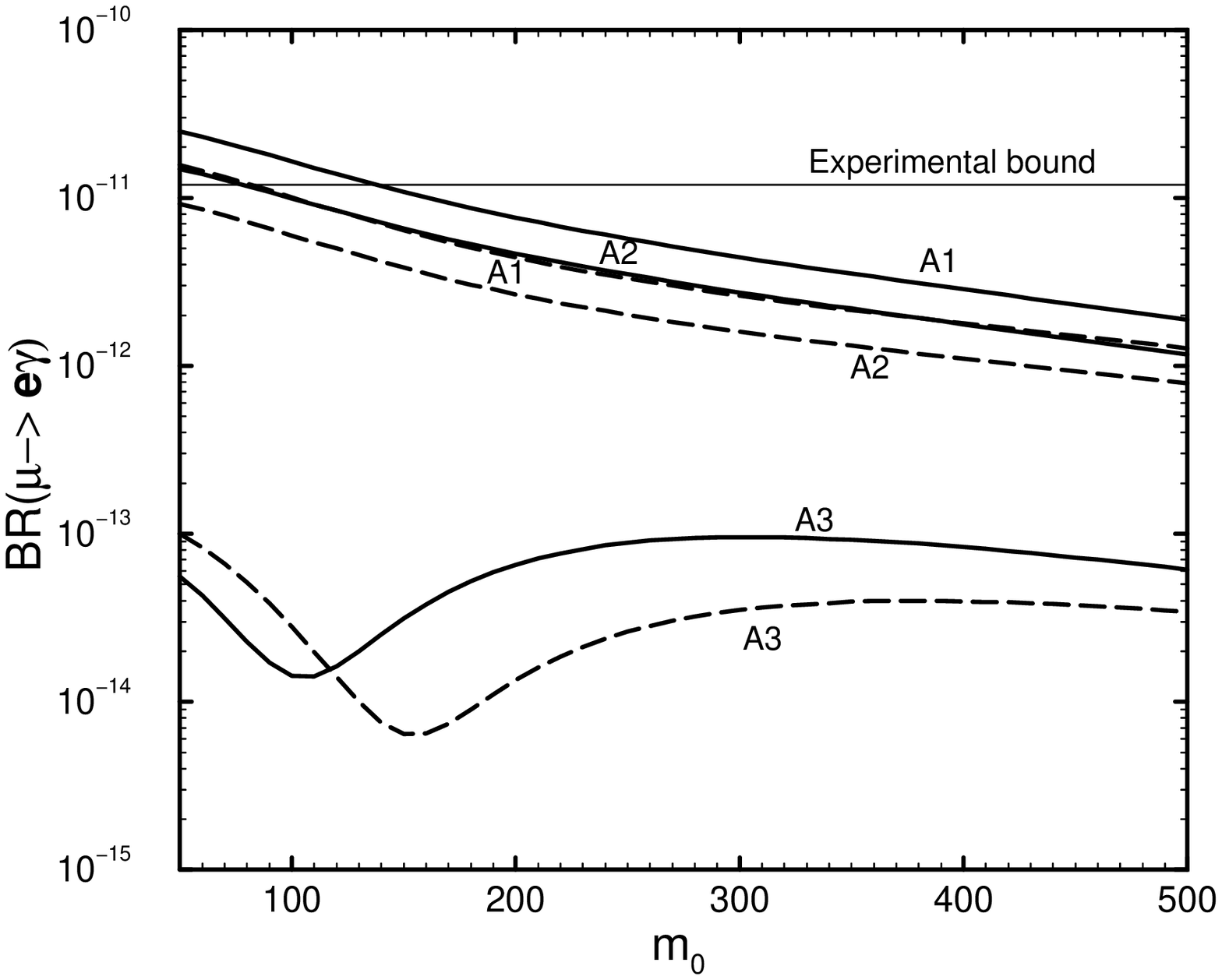,width=3in}
\epsfig{file=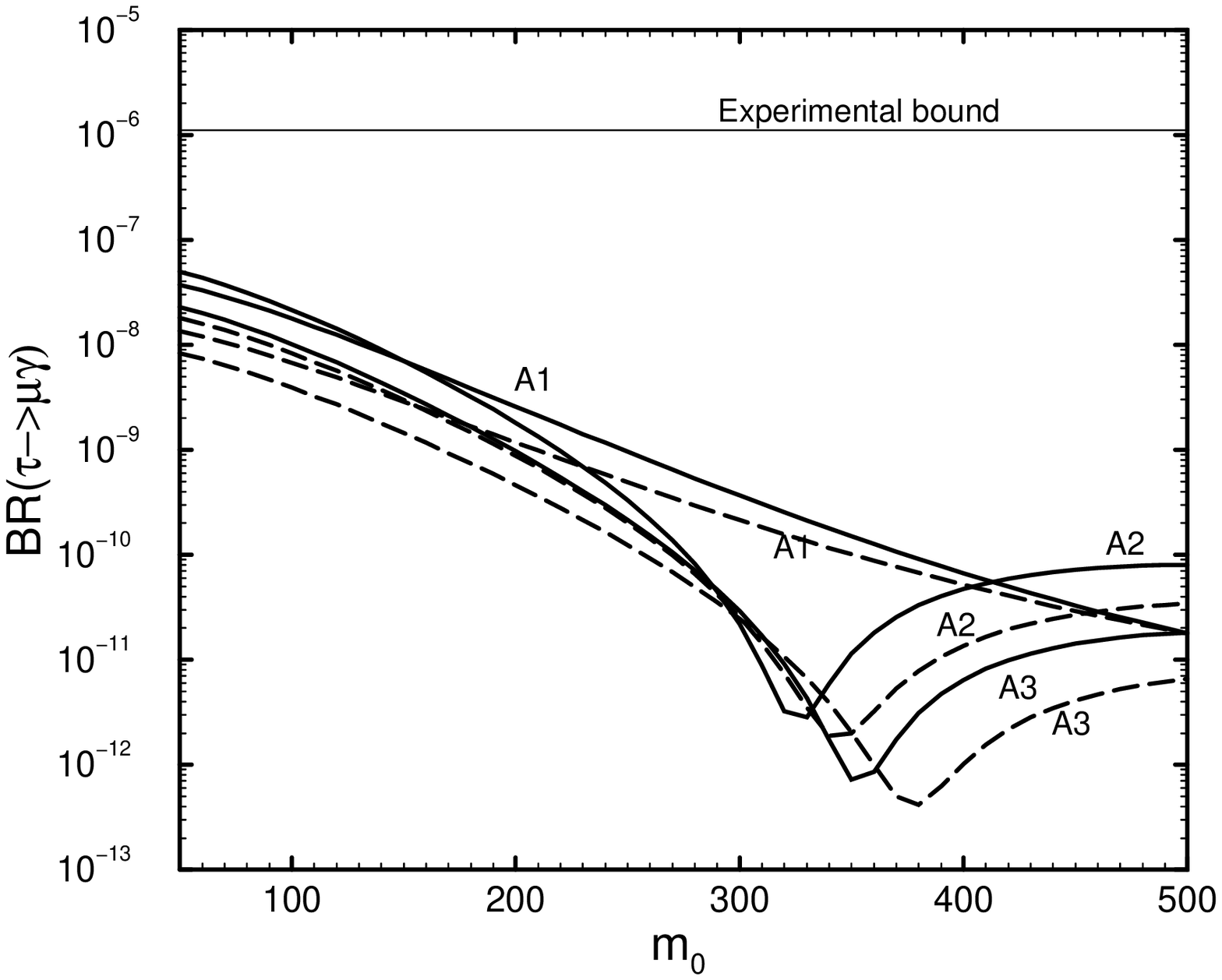,width=3in} \hfill
\end{minipage}
\caption{\it Predictions for $BR(\mu \ra e \gamma)$ and $BR(\tau \ra \mu
\gamma)$ for  texture A   of Section 4, assuming the values
$m_{1/2}=250$ GeV,  $\tan\beta=3$ and  $A_0=-m_{1/2}$. The
solid lines correspond to positive $\mu$, and the dashed ones to
$\mu<0$. The results are for the three specific choice of the 
undetermined numerical coefficients $c_{ij}$ shown in the text.
We see that, for fixed $m_0$, the $\mu\ra e\gamma$ curves are more
sensitive to the 
$c_{ij}$ than are those for $\tau\ra \mu\gamma$.}
\label{cl2}
\end{figure}

\begin{figure}
\begin{minipage}[b]{8in}
\epsfig{file=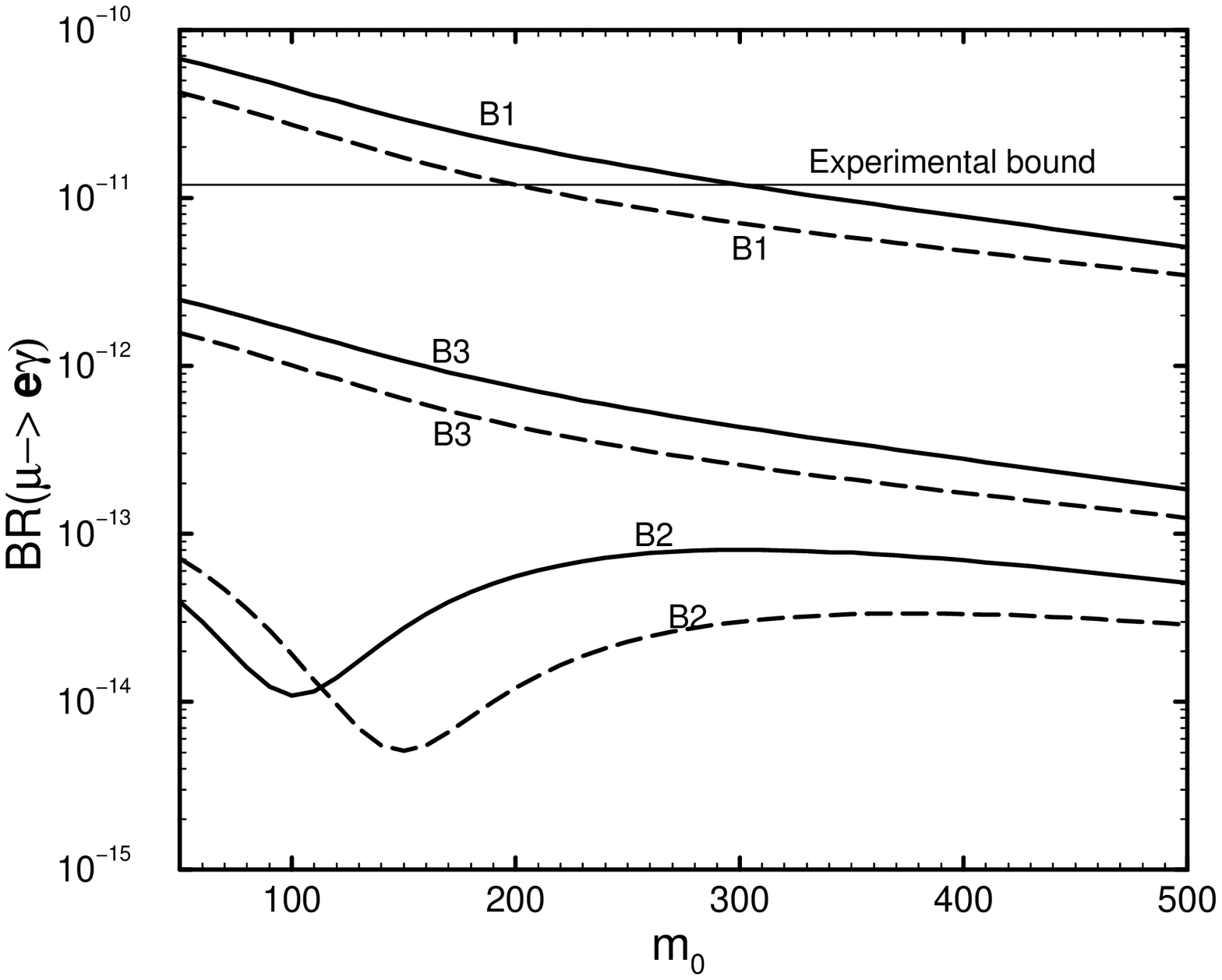,width=3in}
\epsfig{file=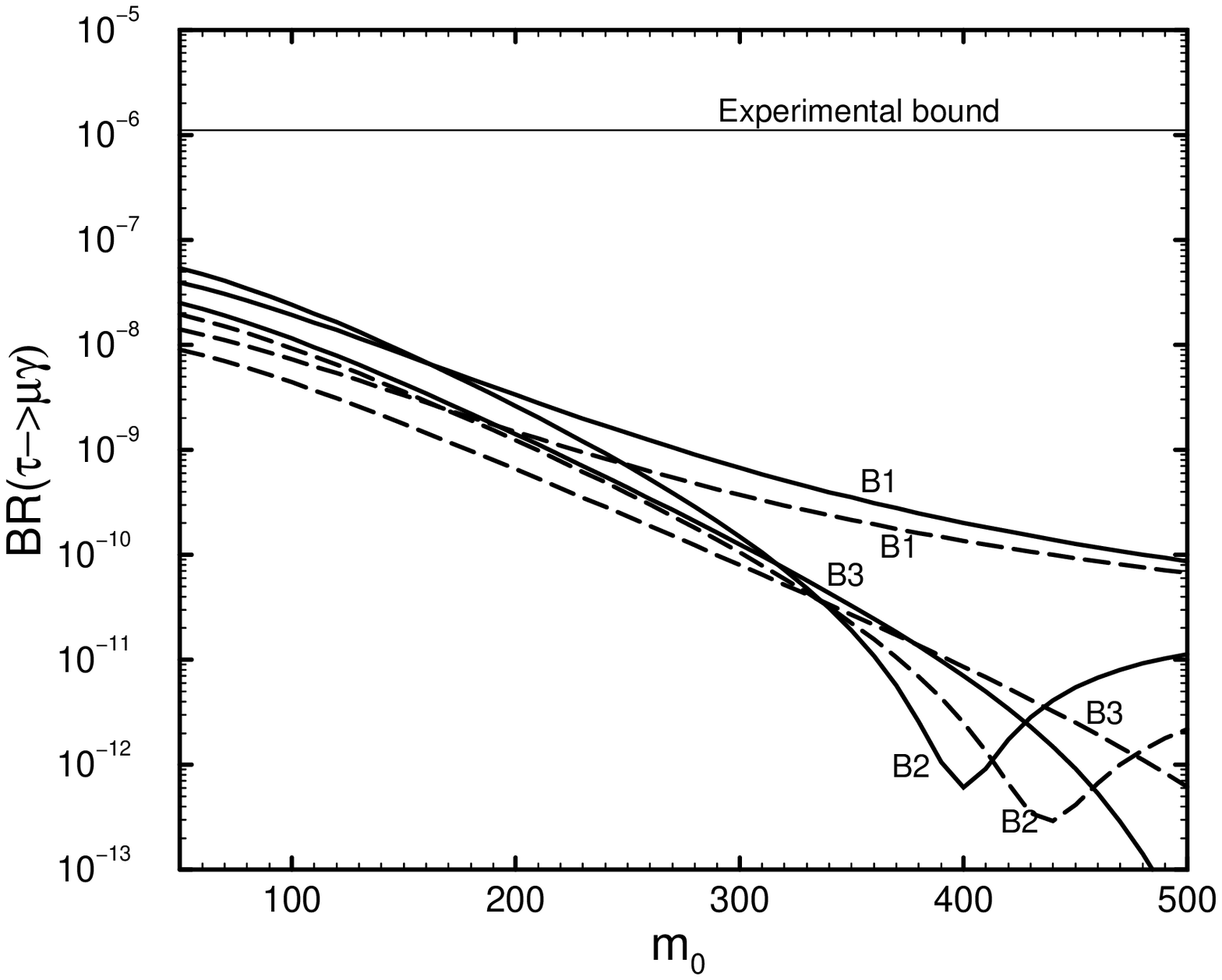,width=3in} \hfill
\end{minipage}
\caption{\it As in Fig.~1, but for texture B of Section 4.}
\label{cl3}
\end{figure}

\begin{figure}
\begin{minipage}[b]{8in}
\epsfig{file=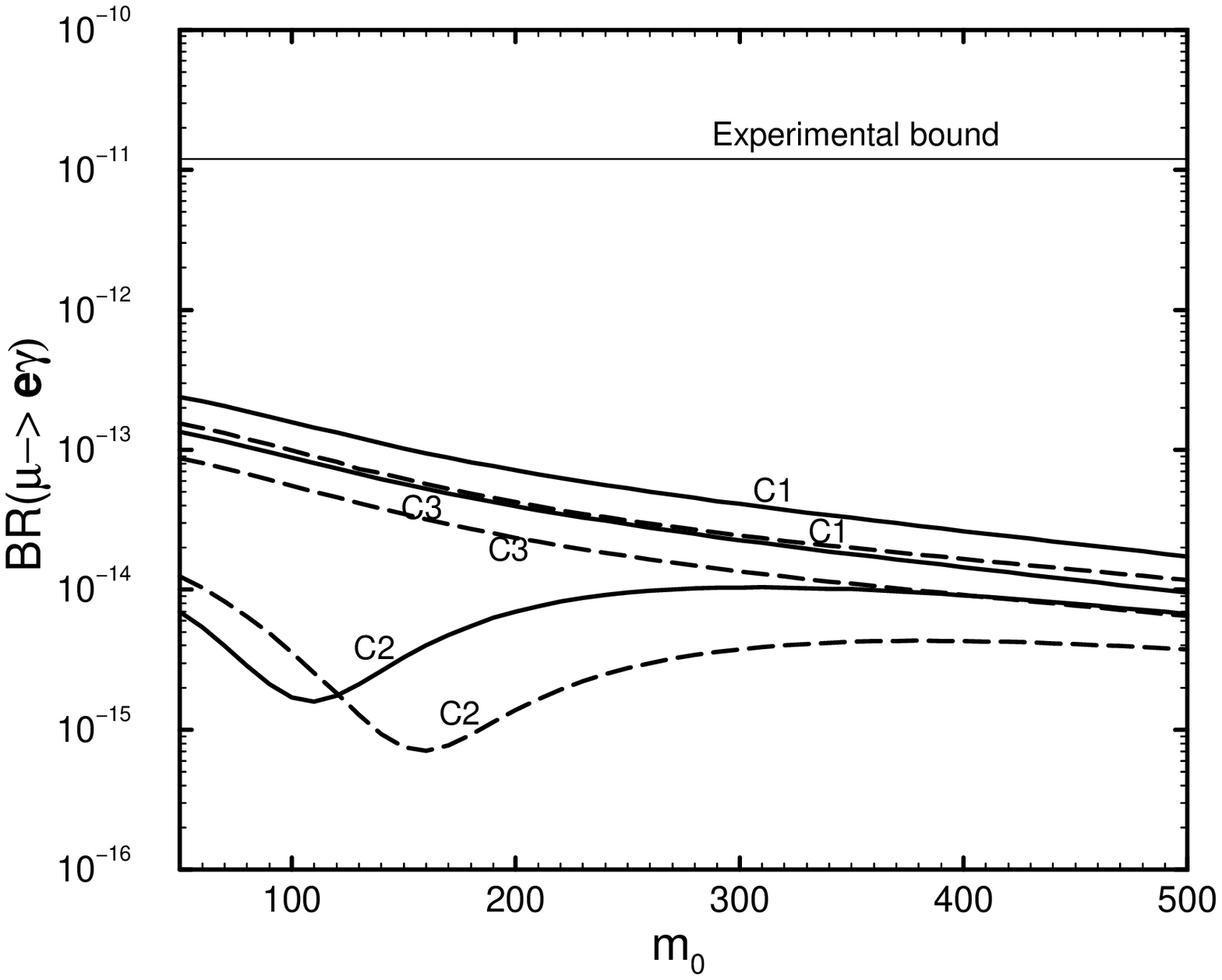,width=3in}
\epsfig{file=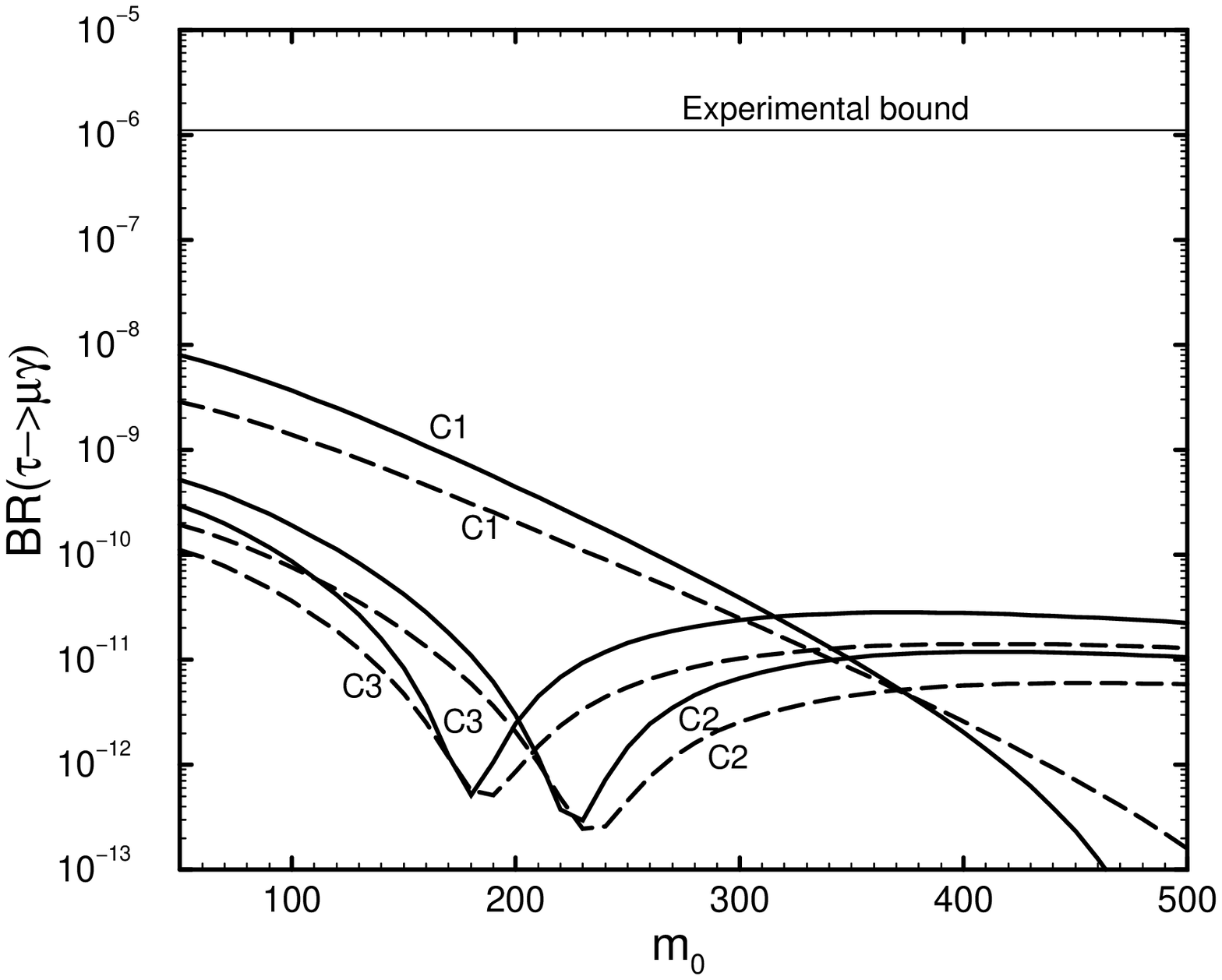,width=3in} \hfill
\end{minipage}
\caption{\it As in Fig.~1, but for texture C of Section 4.}
\label{cl4}
\end{figure}

Among the three cases studied, texture B generally gives
the highest predictions for $BR(\mu\ra e\gamma)$. This is, of course, a
simple consequence of the fact that the (1,3) mixing matrix element
is larger in case B than in case A. Texture C, on the other hand, 
predicts $BR(\mu\ra e\gamma)<10^{-12}$ for all $c_{ij}$ choices. 
As for $BR(\tau\ra \mu \gamma)$, we note that it is in general
enhanced for small $m_{0}$ values. Also,
texture A may yield $BR(\tau\ra\mu
\gamma$) as high as $10^{-7}$.

We present predictions for the larger value
$\tan\beta = 10$ in Fig.~5, for both the $\mu\ra e\gamma$
and $\tau\ra \mu \gamma$ branching ratios, using the three sets
of coefficients $c_{ij}$ for texture A. 
Again, the plots are
obtained using the representative value
$m_{1/2}=250$ GeV and the two possible signs of $\mu$-parameter.
The results are qualitatively similar to those for the
$\tan\beta = 3$ case in Fig.~2, though
the $\mu\ra e\gamma$ branching ratio now tends to lie
above or close to the present experimental limit
for two choices of the coefficients $c_{ij}$.
Only in the third choice of $c_{ij}$ is $BR(\mu\ra
e\gamma$ well below the present experimental limits. 
We also observe in the right-hand plot of Fig.~5 that  the branching
ratios for $\tau\ra \mu \gamma$ are 
enhanced compared to the low $\tan\beta$ case.

\begin{figure}
\vspace*{-0.5in}
\begin{minipage}[b]{8in}
\epsfig{file=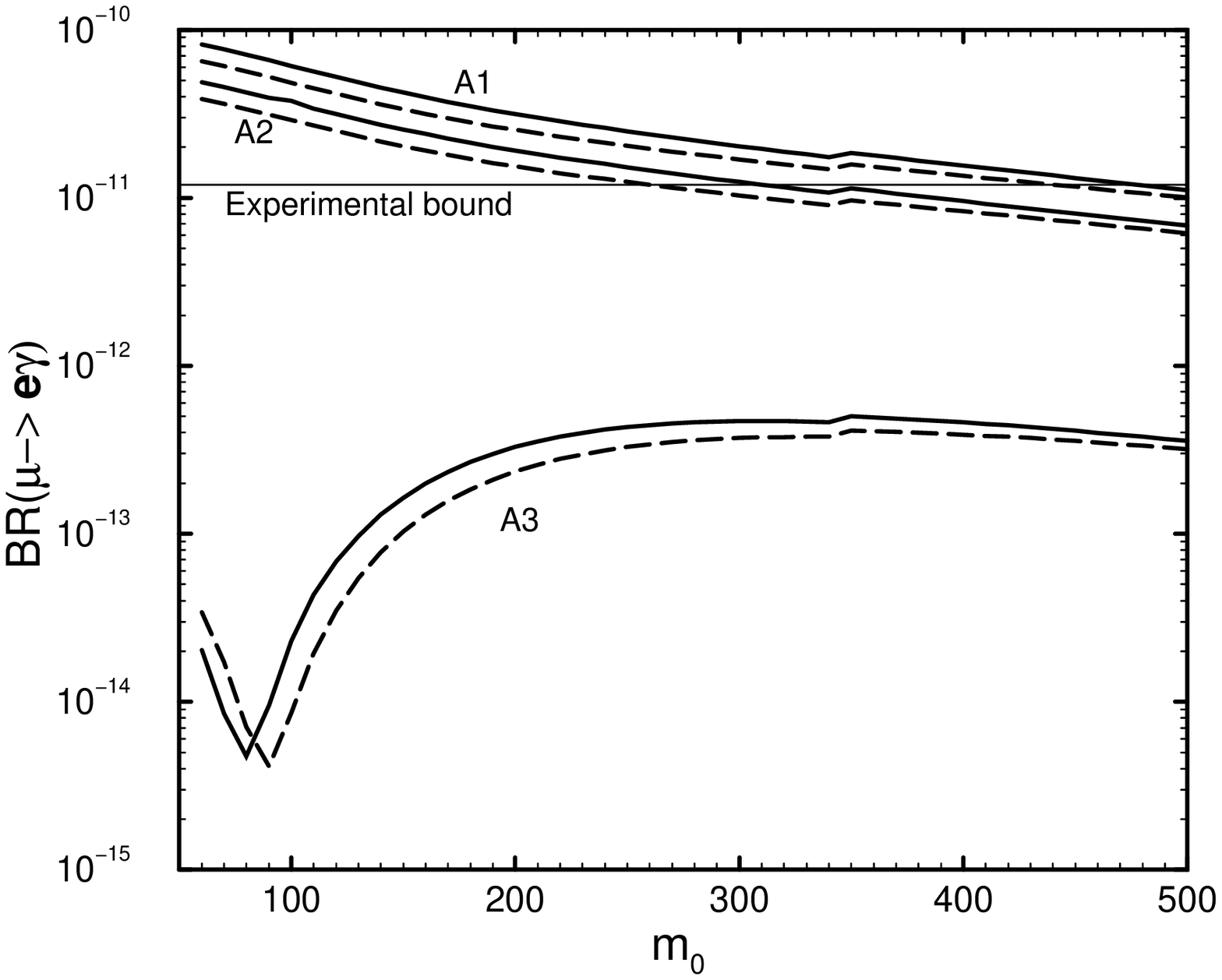,width=3in}
\epsfig{file=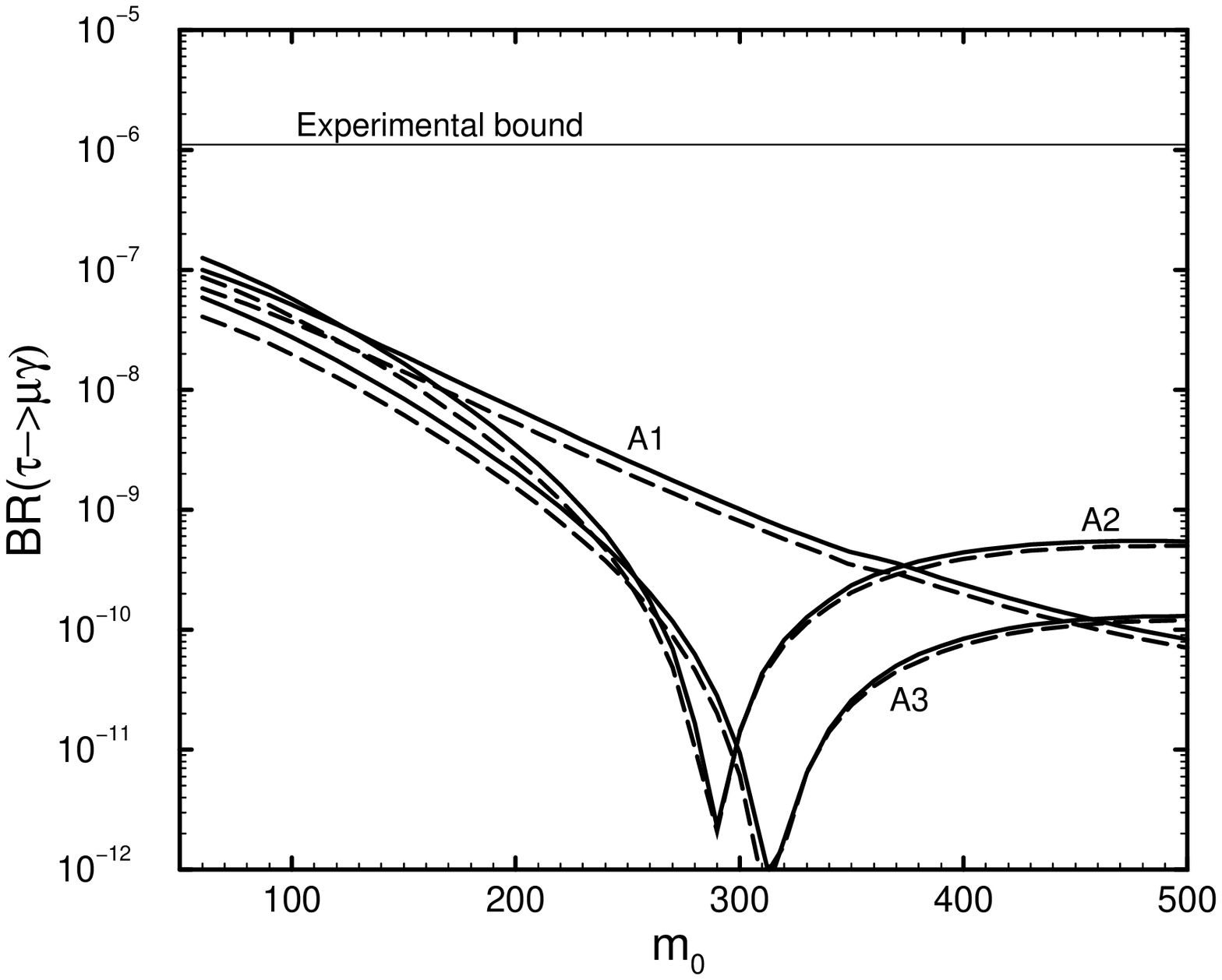,width=3in} \hfill
\end{minipage}
\caption{\it As in Fig.~1, but for $\tan\beta = 10$.}
\label{cl5}
\end{figure}

\begin{figure}
\begin{minipage}[b]{8in}
\epsfig{file=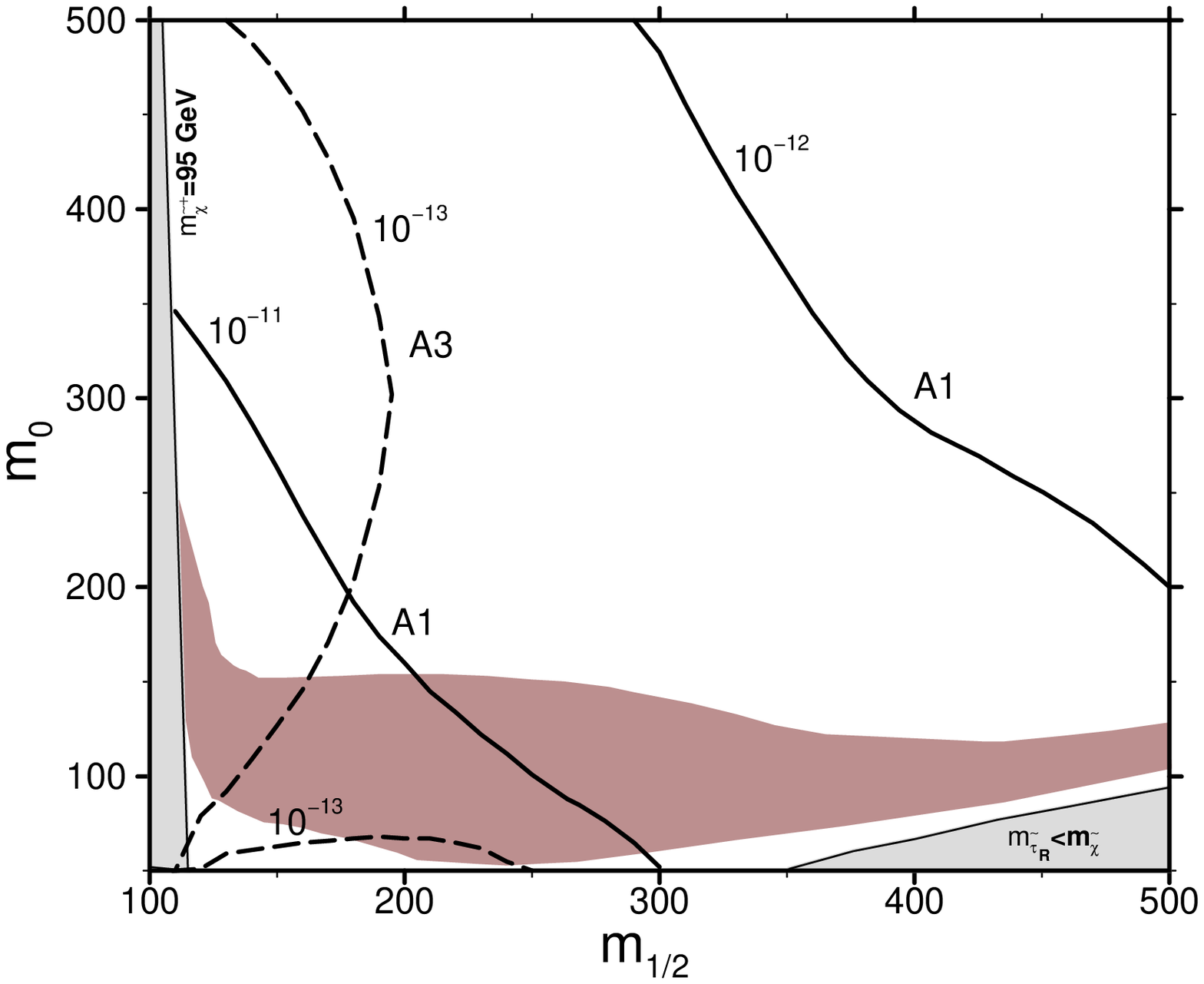,width=3in}
\epsfig{file=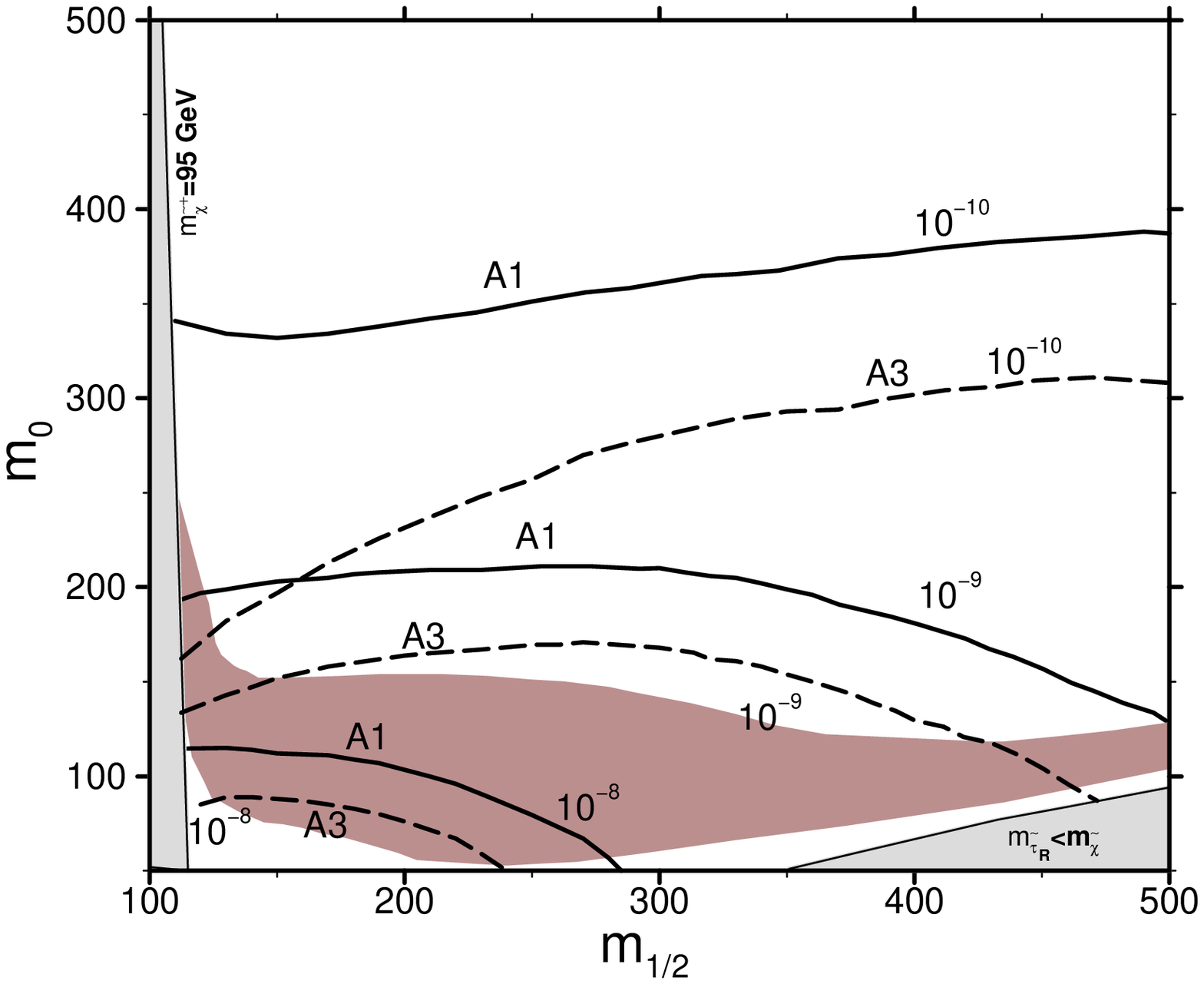,width=3in} \hfill
\end{minipage}
\caption{\it
Contour plots in the $(m_{1/2}, m_0)$ plane for the
decays $\mu \ra e \gamma$ (left) and $\tau\ra \mu \gamma$ (right),
assuming $\tan\beta = 3$ and $\mu < 0$, for the cases $A_1, A_3$.
We see that the rates for both decays are encouraging throughout
the dark-shaded region preferred by astrophysics and cosmology~\cite{ES}.}
\label{metan3}
\end{figure}

\begin{figure}
\begin{minipage}[b]{8in}
\epsfig{file=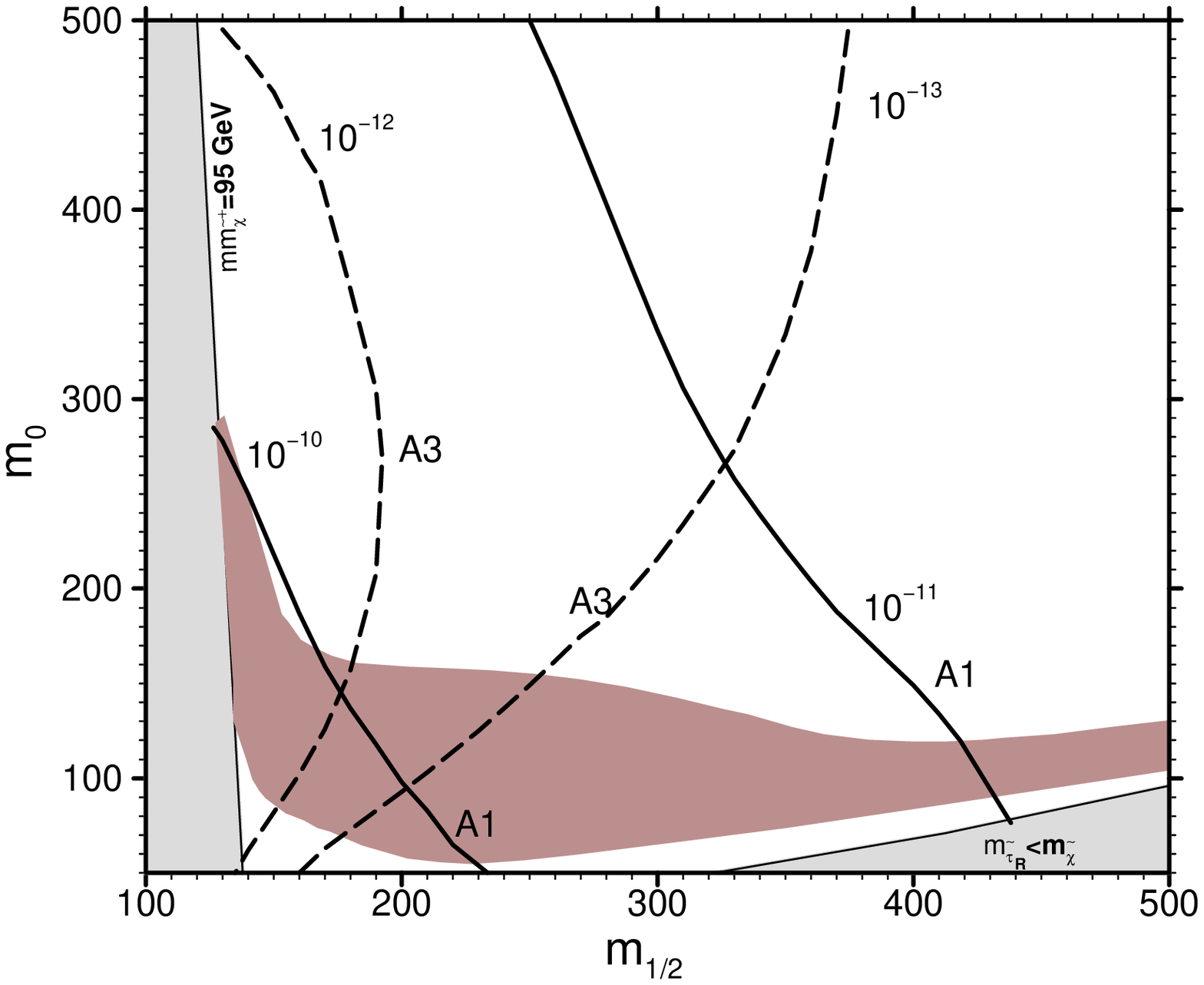,width=3in}
\epsfig{file=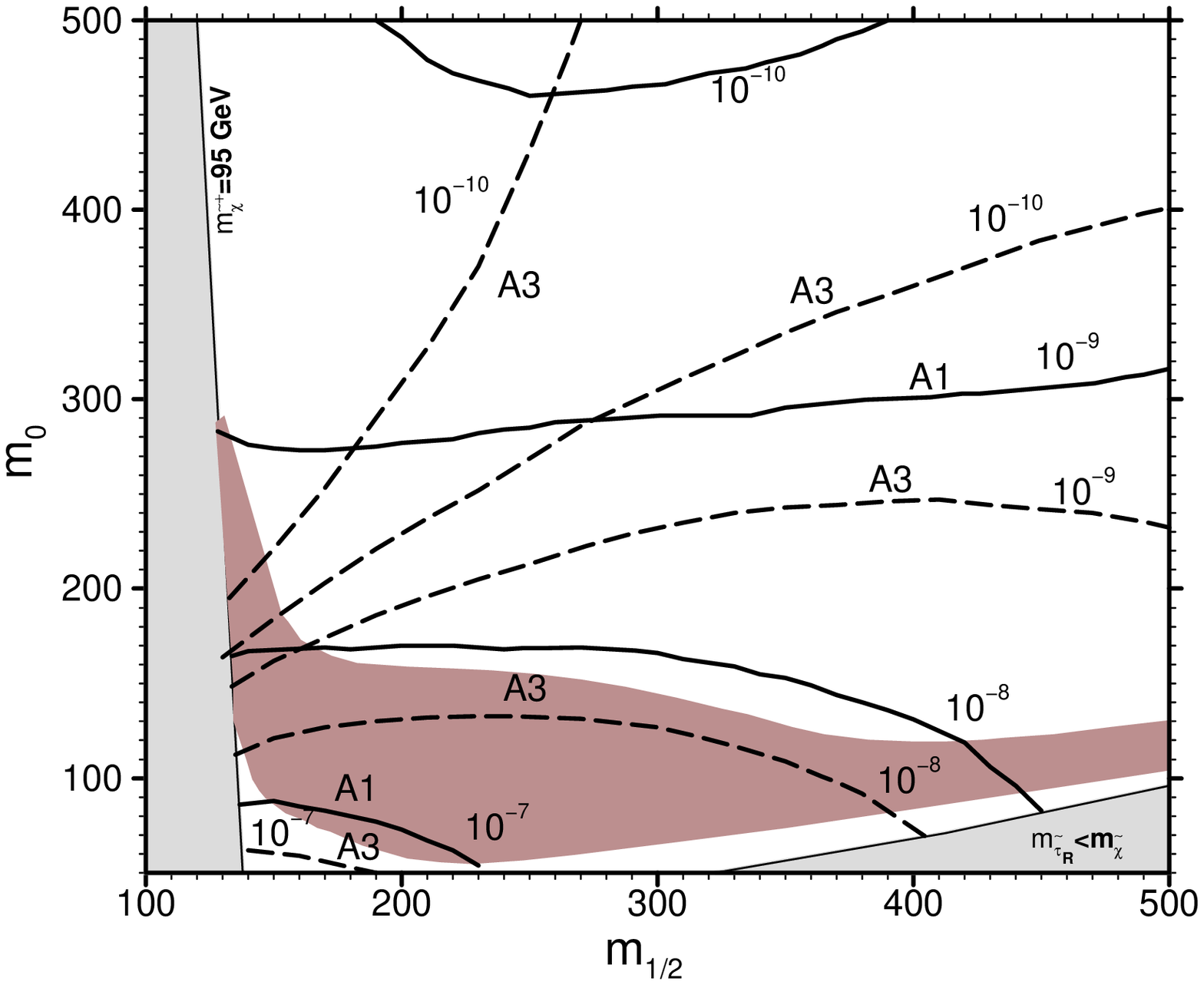,width=3in} \hfill
\end{minipage}
\caption{\it
Contour plots in the $(m_{1/2}, m_0)$ plane for the
decays $\mu \ra e \gamma$ (left) and $\tau\ra \mu \gamma$ (right),
assuming $\tan\beta = 10$ and $\mu < 0$,
and using the cases $A_1$(solid lines),
$A_3$ (dashed).
The rates for both decays are enhanced relative to the small $\tan\beta$ case.
As a result, case $A_3$ is more favourable now in
the dark-shaded region preferred by astrophysics and cosmology~\cite{ES}.}
\label{metan10}
\end{figure}

We show in Figs.~6,~\ref{metan10} predictions
of the branching ratios for the radiative decays
$\mu \ra e \gamma, \tau \ra \mu \gamma$
in the $(m_0, m_{1/2})$ plane, specializing to texture A,
using $\tan\beta=3$ and $10$ and assuming $\mu <0$.
In general, the branching ratios tend to decrease
as $m_{1/2}$ increases. If $\tan\beta=3$ as shown in Fig.~\ref{metan3},
case $A_1$ predicts values of $BR(\mu\ra e\gamma)$ compatible with the
experimental
bound in most of the cosmologically preferred region. 
In contrast, if $\tan\beta=10$ as shown in Fig.~\ref{metan10},
acceptable $BR(\mu\ra e\gamma)$ rates are found only
for large values of $m_0 \ge 400 $ GeV). In this latter case,
the $A_3$ choice of coefficients is more favoured.
We do not display similar plots for
the other two textures, but note that texture B is more sensitive
to the choice of coefficients $c_{ij}$,
and that for certain choices of them it
predicts values close to the experimental bounds for a large portion of
the plane $m_0, m_{1/2} \le 500$~GeV.

The light-shaded areas in Fig.~\ref{metan3} correspond to the 
regions of the
$(m_{1/2}, m_0)$ plane that are excluded by LEP searches for
charginos and by the requirement that the lightest
supersymmetric particle not be charged~\cite{ES}. The dark-shaded
areas in Fig.~\ref{metan3} are those where the cosmological relic density
is in the range preferred by astrophysics. We see that both the
decay modes $\mu \ra e \gamma$ and $\tau \ra \mu \gamma$ could
well be measurable throughout this astrophysical region.
In particular, we find $BR(\tau\ra \mu \gamma)>10^{-9}$ in most of this
region, and for
$m_{1/2} < 220$ GeV and $m_0 <110$ GeV there are regions where
$BR(\tau\ra \mu \gamma)>10^{-8}$. These observations also apply
to texture B, and to the choice $\tan\beta = 10$ (not shown).
The predictions
of texture C may also reach above $10^{-9}$, but reach above $10^{-8}$
only
in a small portion of the cosmologically-favoured region. 
Texture B leads to the
highest predictions for $BR(\mu\ra e \gamma)$, 
typically above $10^{-12}$ for most of the
cosmologically-favoured
region of the $(m_0, m_{1/2})$ plane, 
and even reaching values above $10^{-11}$ in a small
portion of the parameter space.
The sensitivity of the branching ratios to different choices of
coefficients $c_{ij}$ can be seen by comparing the solid and dashed
lines in Fig.~\ref{metan3}.
For
$\tan\beta=3$, the obtained values for the branching ratios under consideration
are smaller.  A  considerable sector of the  cosmologically-favoured
region, leads,
however, to values for the branching ratios of the 
same order as the ones discussed above.

We show in Fig.~8 the correlation between
the $\mu \ra e \gamma$ and $\tau \ra \mu \gamma$ branching ratios
for two different choices of
coefficients $c_{ij}$ for each of the textures A, B and C presented in
Section 4. In each case, the branching ratios have been calculated for 
a sampling of $(m_0, m_{1/2})$ pairs
in the cosmologically-favoured region of~\cite{ES} for
$\tan\beta=10$: similar results hold for $\tan\beta = 3$.
We see clearly how
the correlations between the two branching ratios vary with the choices of
the $c_{ij}$ coefficients, because of their influences on flavor mixing.
In the two cases $A_1$ and $A_3$ (texture $A_2$ is similar to $A_1$)
presented in Fig.~8,
the values of the predicted branching ratios
are characteristic for each $c_{ij}$ set, with the case $A_3$  
generally predicting smaller ratios for $BR(\mu \ra e \gamma) /
BR(\tau \ra \mu \gamma)$.
In the case of
texture C, the dependence of the results on the choice of $c_{ij}$
coefficients is rather different: this
texture tends to predict a
relatively large ratio $BR(\mu \ra e \gamma) / BR(\tau \ra \mu \gamma)$.

\begin{figure}[h]
\begin{center}
\epsfig{file=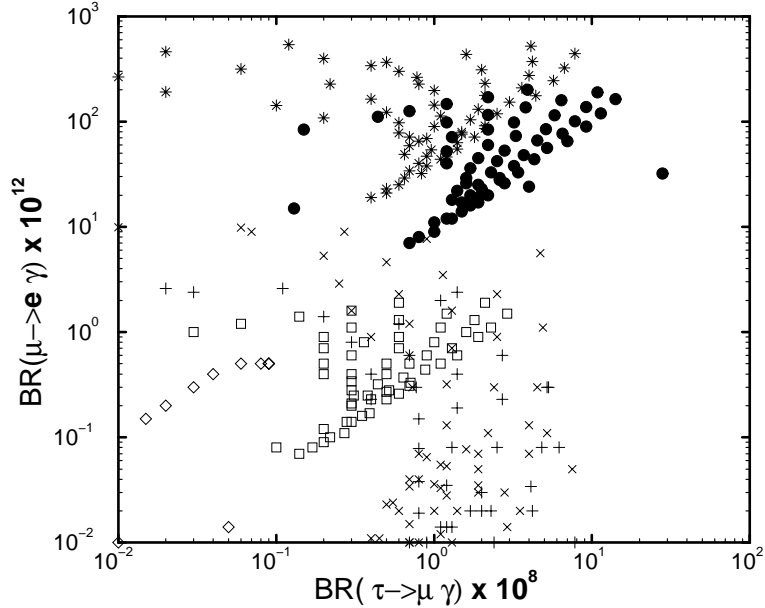,width=10cm}
\end{center}
\caption{\it Scatter plot of model predictions for $BR(\mu\ra e\gamma)$
versus $BR(\tau\ra \mu\gamma)$, assuming 
$tan\beta=10$ and $\mu<0$. The
circles correspond to texture $A_1$,
the {\bf +} to texture $A_2$, 
the stars  to texture $B_1$,
the $\times$ to texture $B_2$
the squares to texture $C_1$,
the diamonds to texture $C_2$.
Note the characteristic correlations in the different models.} 
\label{scp}
\end{figure}

Finally, we show results for $\mu-e$ conversion, using both penguin
and box diagrams. We gave in
Section 2 an order-of-magnitude
estimate of the branching ratio for this reaction, namely
$BR(\mu Ti\ra e Ti) \approx  5.6\times 10^{-3} BR(\mu\ra e\gamma)$.
However, as we commented there, the two processes exhibit 
different functional dependences on the $A_{M/E}^{L,R}$
functions (indeed, as is well known,  $\mu\ra e \gamma$ does not depend at all
on the parameters $A^{E}_{L,R}$). Further, in the case of  $\mu-e$ conversion
there are additional contributions from
box graphs which further complicate the ratio $BR(\mu Ti\ra e
Ti)/BR(\mu\ra e\gamma)$. In view of the possible improvement of the 
experimental sensitivity to $\mu-e$ conversion, we 
present plots similar to given previously for $\mu\ra e\gamma$.

It is instructive to compare the $\mu-e$ conversion rate with the
corresponding predictions for $BR(\mu \rightarrow
e\gamma)$, to explore the
effects of penguin and box diagrams.
In Fig. ~9 we plot  the ratio of the $\mu\ra e$ conversion and $\mu\ra
e\gamma$  rates versus the scalar mass parameter $m_0$.
We see immediately that the ratio is not constant, though its general
order of magnitude is that estimated in (\ref{convratio}).
The dependence of the penguin contribution on $m_0$ is shown
separately from the combined effects of penguin  and box diagrams,
for the two signs of $\mu$. In the case of $\mu>0$, $BR(\mu \ra e\gamma)$
is relatively  enhanced for large $m_0$ values, whilst the opposite
is true for $\mu<0$.

\begin{figure}[h]
\begin{center}
\begin{minipage}[b]{8in}
\epsfig{file=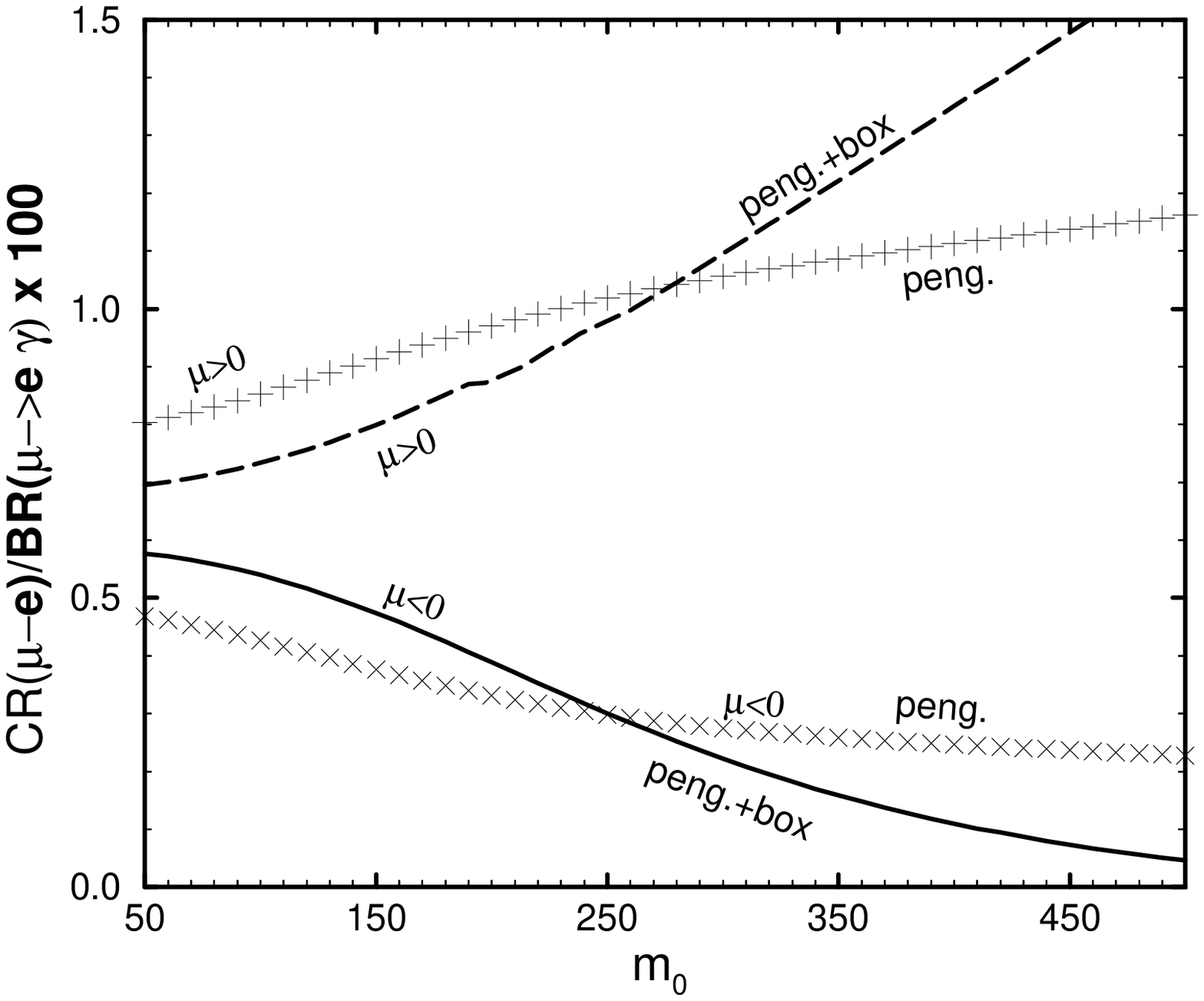,width=3in}
\epsfig{file=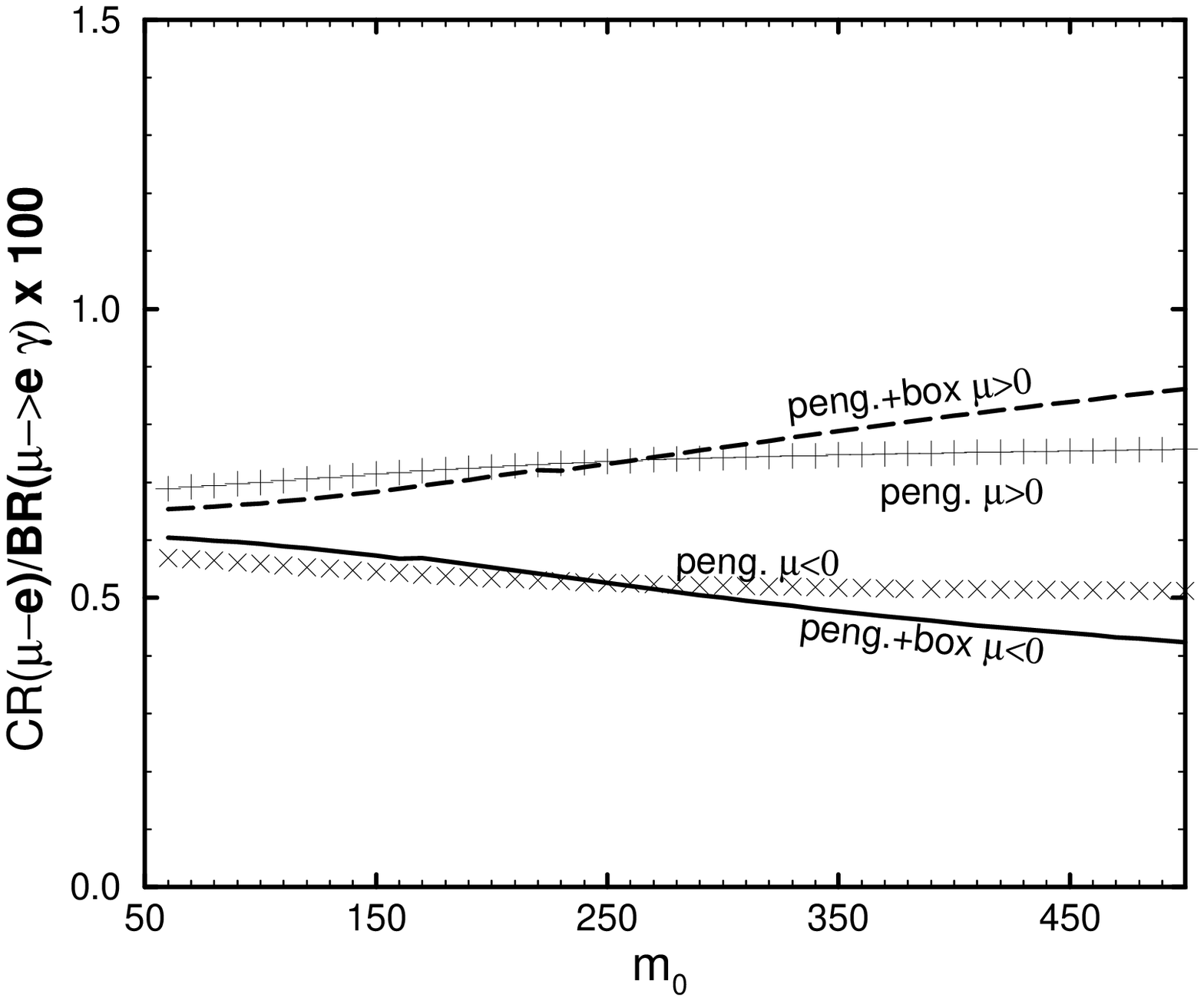,width=3in}
\end{minipage}
\end{center}
\caption{\it The ratio of the $\mu \ra e$ conversion rate to
$BR(\mu \ra e \gamma)$ is plotted versus $m_0$,
for $\mu < 0$. We plot
the penguin contribution separately from the sum of the
penguin and box diagrams, for 
$\tan\beta=3$ (left) and $\tan\beta=10$ (right).
}\label{rat3}
\end{figure}

We show in Fig.~10 the 
dependence of the $\mu \ra e$ conversion rate on $m_0$ for the two
textures $A, B$ and for the values
$\tan\beta=3$ and $m_{1/2}=250$ GeV. As before, we choose three sets
of numerical coefficients $c_{ij}$ for each texture, and exhibit the
results for both
signs of $\mu$.  Interestingly, cases 
$A_1, A_2$ give a rate close to the present experimental limit
for most of the $m_0$ region explored when $\mu>0$. The corresponding
predictions for $\mu<0$ are considerably lower for large $m_0$ values,
but converge with those of the $\mu>0$  case  for small $m_0$.
We note that texture B exhibits greater sensitivity to the coefficients
$c_{ij}$ than does texture A.

\begin{figure}[h]
\begin{minipage}[b]{8in}
\epsfig{file=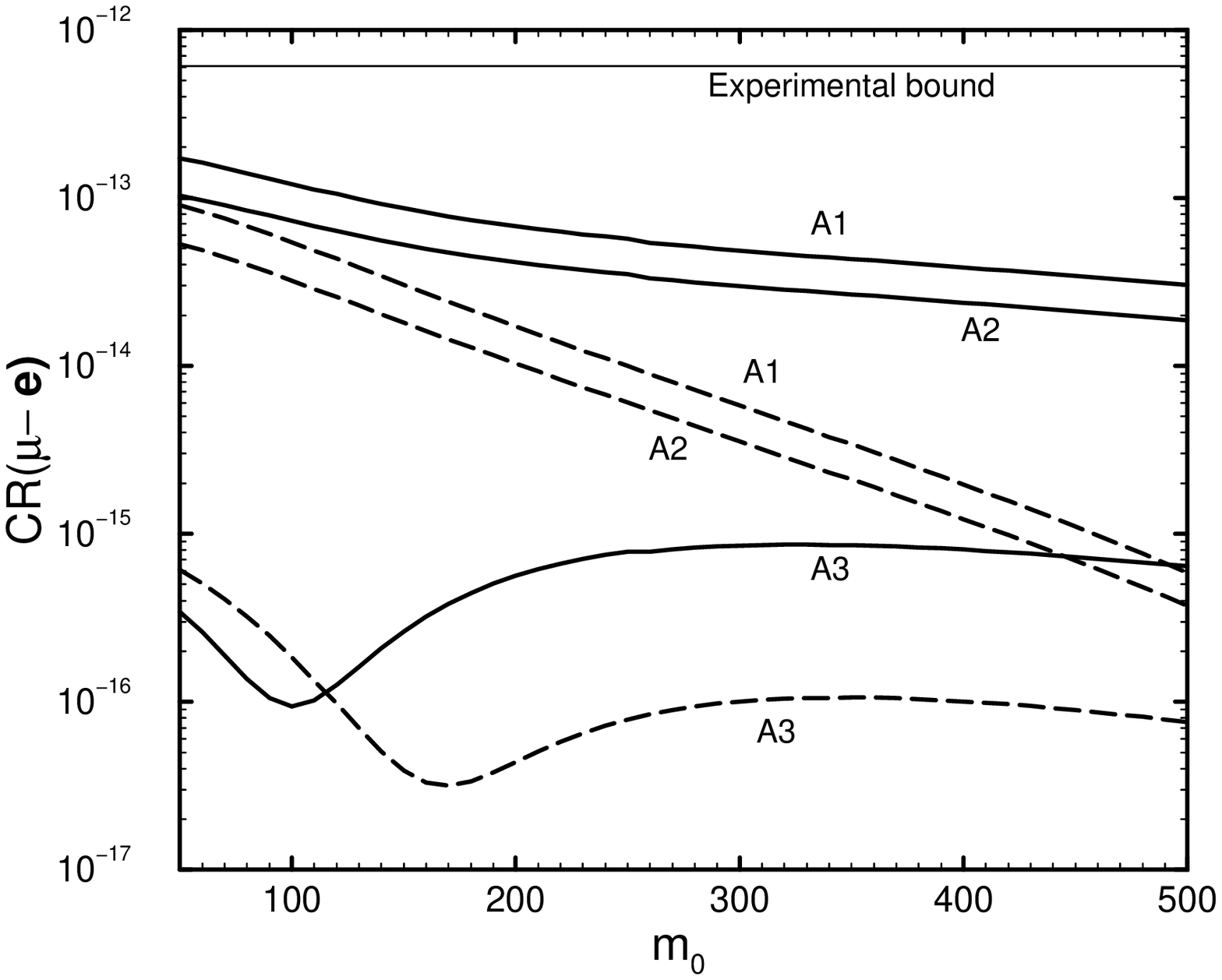,width=3in}
\epsfig{file=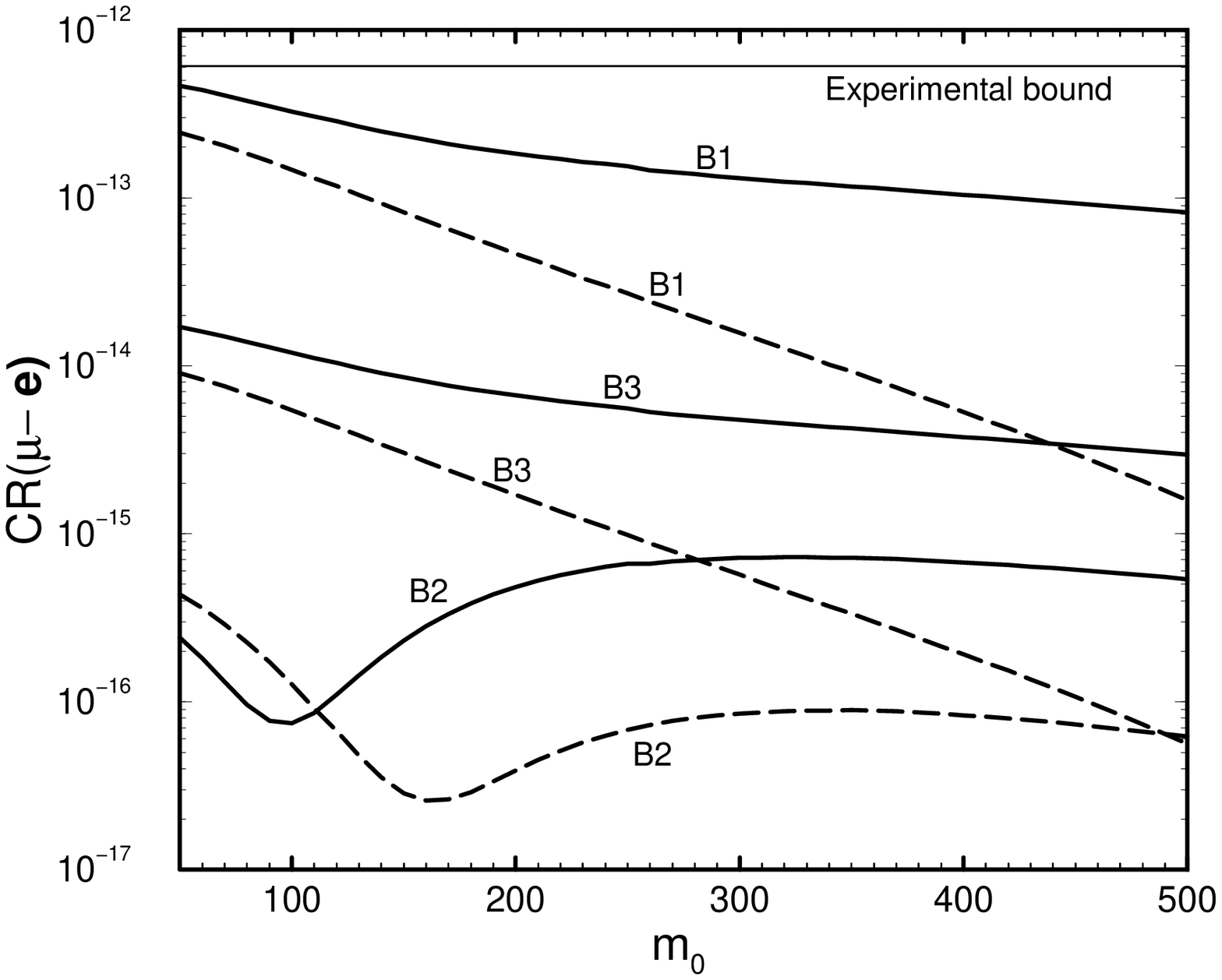,width=3in} \hfill
\end{minipage}
\caption{\it The muon conversion rate $\mu \ra e$ as a
function of $m_0$ for the  textures A and B, assuming
$\tan\beta = 3$. The solid lines correspond to
$\mu > 0$ and the dashed to
$\mu < 0$.}
\label{mecol3}
\end{figure}

\begin{figure}[h]
\begin{minipage}[b]{8in}
\hspace*{2.5 cm}
\epsfig{file=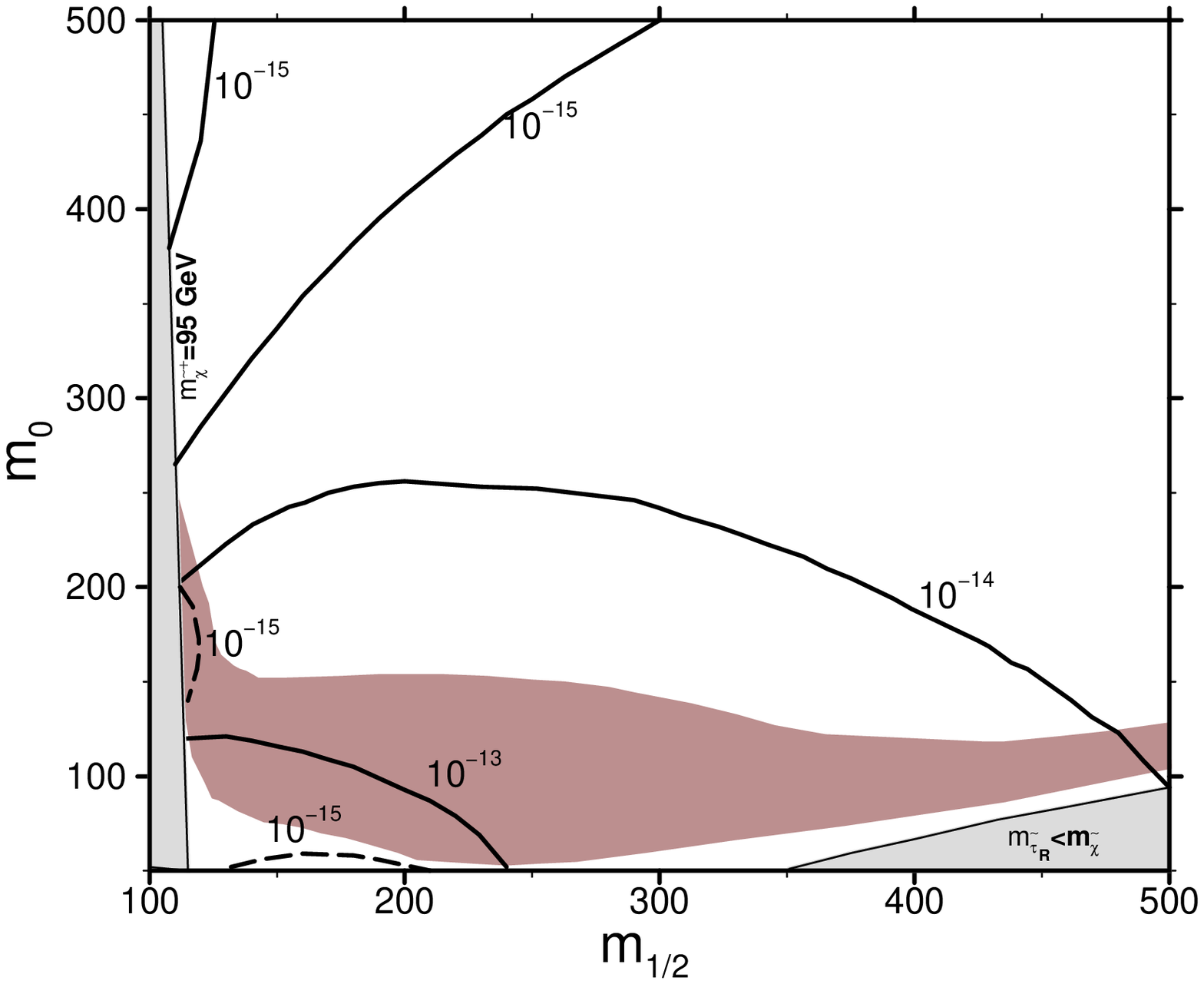,width=10cm}
\end{minipage}
\caption{\it
Contour plots in the $(m_{1/2}, m_0)$ plane for
$\mu \ra e $ conversion,
assuming $\tan\beta = 3$ 
and $\mu < 0$ and using the cases $A_1, A_3$.
We see that the conversion rate is encouraging
throughout the dark-shaded region preferred by astrophysics and
cosmology~\cite{ES} in the scenario $A_1$.}
\label{come}
\end{figure}

Fig.~11 displays contours of the rate for $\mu \ra e$
conversion in the $(m_0, m_{1/2})$ plane for the two scenarios
$A_1$ and $A_3$. We see that the former predicts a rather larger
rate, which offers good prospects for observation throughout
the region preferred by cosmology. On the other hand, scenario
$A_3$ predicts a rather lower rate for $\mu \ra e$ conversion.

We comment at this stage on some of the ambiguities in the
results shown above, within the general $U(1)$ framework
studied. We have already explored to some extent the ambiguity
associated with ${\cal O} (1)$ coefficients in the Dirac
mass matrices for the fermions. The ambiguity induced by
simple sign changes can be particularly acute, as we
illustrate with one simple exercise. For example, since the
light Majorana mixing matrix is $V_{MNS} = V_\ell V_\nu^{\dagger}$,
any modification that changes the relative signs of off-diagonal entries
in $V_\ell$ and $V_\nu$ could cause large changes in the
mixing angles of $V_{MNS}$, as one changes from destructive to
constructive interference, or vice versa, with intermediate
possibilities corresponding to various phase possibilities that
are not specified by the $U(1)$ symmetry.

\begin{figure}[h]
\begin{center}
\epsfig{file=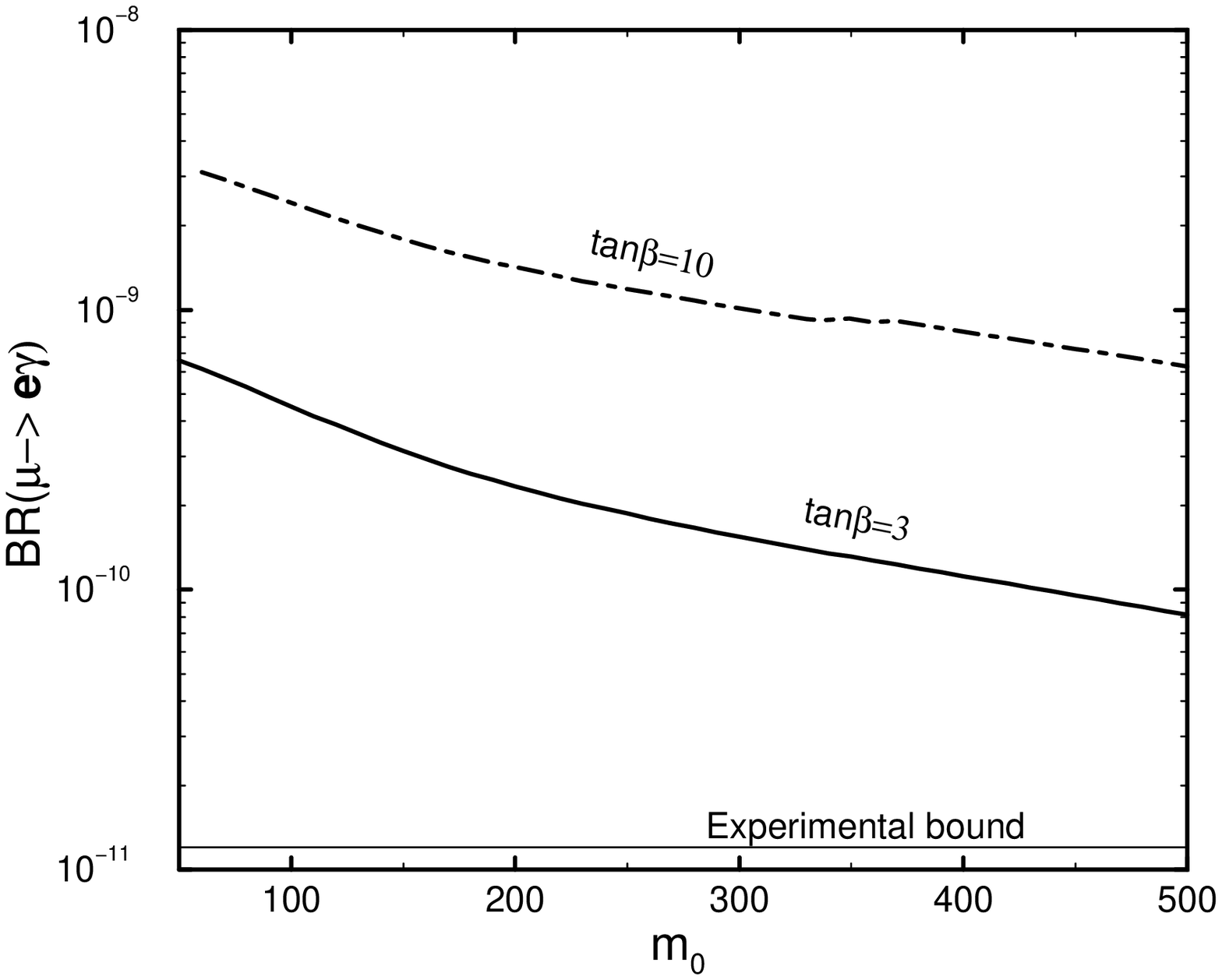,width=3in}
\epsfig{file=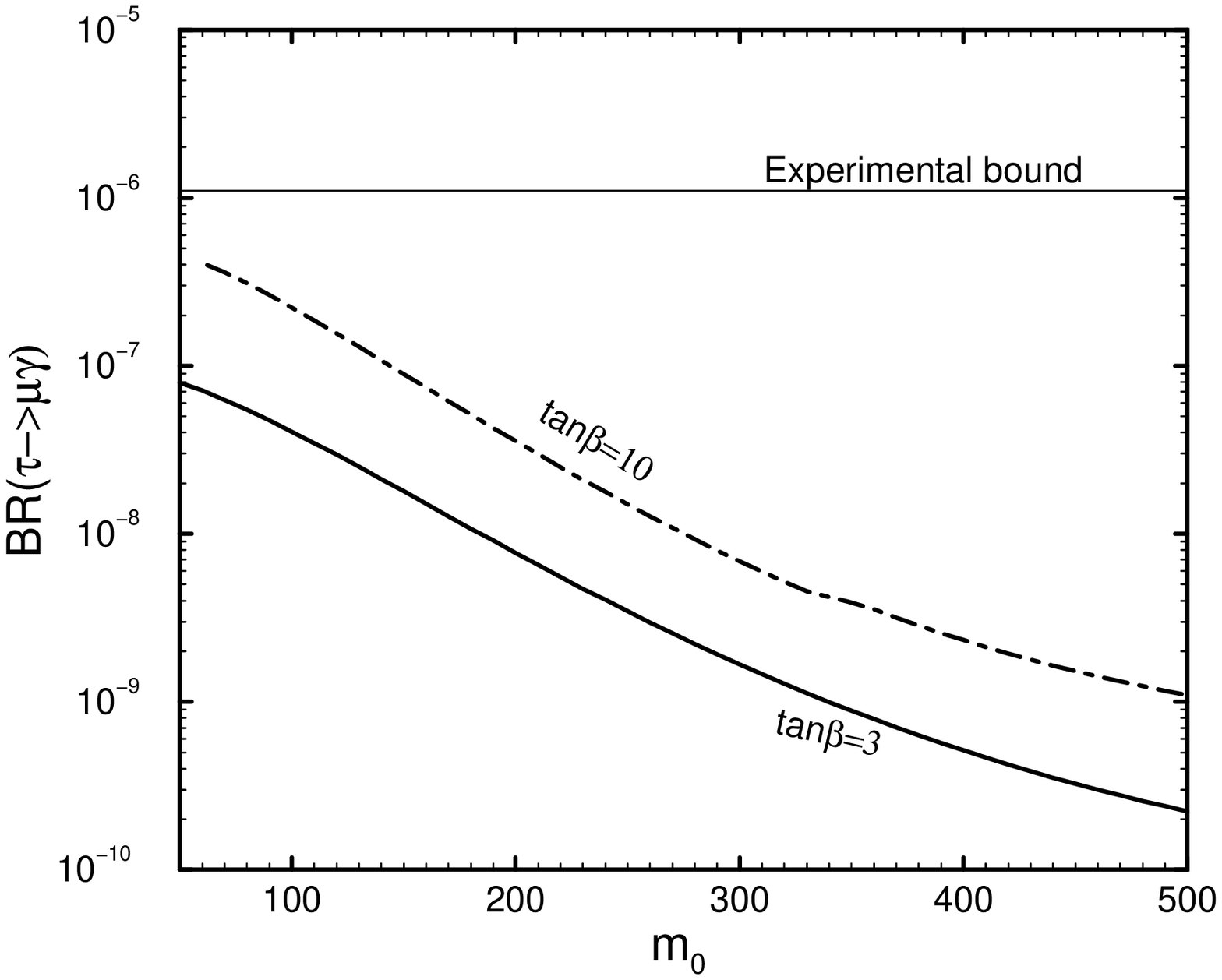,width=3in}
\end{center}
\caption{\it Predictions of $BR(\mu \ra e \gamma)$ and
$BR(\tau \ra \mu \gamma)$ 
in the `inverted' version of
$A_1$, obtained by inverting the signs of all the
off-diagonal terms in $m_\nu^D$, for $\mu < 0$.}
\label{fig:A1lar}
\end{figure}

As an exercise, using the
numerical values of the coefficients in the case $A_1$ discussed above,
we invert the signs of all the off-diagonal entries in $m_{\nu}^D$, and
repeat consistently all the subsequent steps in the calculations.
The implications for $\mu \ra e \gamma$ and $\tau \ra \mu \gamma$ are
shown in Fig.~\ref{fig:A1lar}, where we see that $BR(\mu \ra e \gamma)$
may be increased by about two orders of magnitude~\footnote{We find a
similar enhancement for the $\mu \ra e$ conversion rate.},
whereas that for
$BR(\tau \ra \mu \gamma)$ is increased by less than an order of magnitude.
This `inverted' $U(1)$ model actually has much larger $\nu_\mu - \nu_\tau$
mixing than the `uninverted' version of $A_1$, or indeed the other
textures studied previously, agreeing better with the
Super-Kamiokande data. We interpret the difference between $A_1$ and
`inverted' $A_1$, on the one side, and between $A_1$ and $A_3$, on the
other side, as indicative of the numerical ambiguity within any
particular texture~\footnote{The drastic changes in the `inverted' $A_1$
case may not be possible in GUT models
where the neutrino Dirac matrix is related closely to the up-quark mass
matrix. In such models,
large $\nu_{\mu}-\nu_{\tau}$ may also be arranged by a suitable
choice of the heavy Majorana sector.}. 
We see that the predictions for $\mu \ra e \gamma$
are very variable, bracketing the present experimental upper limit within
a broad range, whereas the predictions for $\tau \ra \mu \gamma$ are
more closely bunched, and likely to be within reach of experiment.

The generality of these features is left for exploration in the future.
There are also other ambiguities, for example in the choice of the
heavy Majorana mass matrix~\cite{ELLN}, whose detailed study
we also leave for future work.

Before concluding, we make a few more remarks about 
non-Abelian models, and about the flipped $SU(5)$
model. As commented at the end of subsection IV-B, 
generic non-Abelian models would fall into our `mismatched' category,
and we would expect them to have relatively large rates
for both $\mu \ra e$ and $\tau \ra \mu$ transitions, as a
result of their preference for near-bi-maximal mixing.
Therefore, we expect the prospects for charged-lepton-flavour
violation in these models to be at least as favourable as in
the Abelian models studied here. In the case of flipped $SU(5)$,
if we use naively the
matrices displayed in Section 5, we find that the $\mu
\ra e \gamma$ process is rather enhanced, and 
it exceeds the present experimental bounds in a considerable
region of the parameter space, at least for some generic
choices of undetermined numerical coefficients. However,
since the plethora of
poorly-constrained expansion parameters introduces
ambiguities, as discussed earlier,
a complete exploration of this model is beyond the scope of
this paper, and it may be that the model can survive for suitable values
of these coefficients.
On the other hand, the $\tau \ra \mu \gamma$ reaction is highly
suppressed,
because in this model all the mixing needed to interpret the atmospheric
neutrino data
comes from the neutrino sector, and there is no mixing
in the $\tau-\mu$ charged-lepton sector.

\section{Conclusions}

Although family symmetries provide many interesting insights
into the hierarchy of fermion masses, there is no unique
framework that fits the available information on charged
fermions and neutrinos. As a result, there is considerable
ambiguity, even within the subclass of Abelian flavour models,
in their predictions for charged-lepton-flavour violation.
We have explored some of the range of possibilities in this paper.
Many sets of undetermined ${\cal O}(1)$ 
coefficients in Abelian flavour symmetry models of the
charged lepton mass matrix can fit well the mass spectrum and
the neutrino data, but vary in their predictions
for the branching ratios of rare processes. As examples,
three different sets of coefficients were used to fit the charged lepton
and quark mass matrices in each of three Abelian texture models.
Our studies show that these vary
in their predictions for flavour-changing branching ratios by up
to two orders of magnitude. The good news, however, is that many of
these models seem to be accessible to a new round of experiments
searching for $\mu \ra e$ and $\tau \ra \mu$ transitions.
In particular, we would like to re-emphasize the interest of
exploring the branching ratio for $\tau \ra \mu \gamma$ down to
the $10^{-9}$ level or below, as may be possible at the LHC.

In view of their accessibility, and precisely because of
their model-dependence, such rare decays may become a
powerful tool for distinguishing between different
neutrino textures. As we have seen,
textures based on Abelian
flavour symmetries tend to predict relatively small $\mu-e$ 
flavour mixing, thus leading to rates for
$\mu \rightarrow e \gamma$,
$\mu \ra 3e$ and $\mu \to e$ conversion in nuclei
that are generically below the experimental bounds,
though close enough to offer interesting physics
opportunities for experiments with present and future
intense $\mu$ sources. On the other hand, schemes based on
non-Abelian flavour symmetries would tend to predict
large mixing in the (1-2) lepton sector.
In this case, larger rates are likely to be found for the
above processes, and in certain textures
part of the supersymmetric parameter space may already be excluded.
It is likely also that string-derived flipped $SU(5)$ schemes based on
would have large off-diagonal entries
in the (1-2) lepton sector, leading to larger rates
for $\mu \rightarrow e \gamma$,
$\mu \ra 3e$ and $\mu \to e$ conversion in nuclei than
the simplest models with Abelian flavour symmetries,
though the rates for $\tau \ra \mu \gamma$ would be relatively low.

We conclude by encouraging the community to pursue actively
new generations of experiments to probe charged-lepton-flavour
violation, in both $\mu$ and $\tau$ decays. Such efforts would
complement nicely the physics being revealed by neutrino oscillations,
and could provide precious insight into the flavour problem.

\noindent
{\bf Acknowledgements}\\[2.5mm]
M.E.G. thanks D. Carvalho for useful discussions, and his research
has been supported by the European Union TMR Network contract
ERBFMRX-CT96-0090.

\noindent
{\bf Note Added}\\[2.5mm]
While this paper was in the final stages of preparation, we received
the paper by Feng, Nir and Shadmi in~\cite{FNS}, which makes
points similar to ours, in a complimentary way.

\end{document}